\documentclass[fleqn,usenatbib]{mnras}

\usepackage{newtxtext,newtxmath}
\usepackage{mathtools}
\usepackage{amsmath}

\usepackage{soul}
\usepackage{subfig}
\usepackage{enumitem}

\usepackage{threeparttable}
\usepackage{booktabs}
\usepackage{multirow} 

\usepackage[T1]{fontenc}
\usepackage{comment}
\DeclareRobustCommand{\VAN}[3]{#2}
\let\VANthebibliography\thebibliography
\def\thebibliography{\DeclareRobustCommand{\VAN}[3]{##3}\VANthebibliography}

\usepackage{graphicx}	
\usepackage{amsmath}	
\usepackage{tabularx}
\usepackage[colorinlistoftodos,color=violet!30!white,size=scriptsize]{todonotes}
\usepackage{xcolor}

\usepackage{color}
\usepackage{xcolor}

\newcommand{\OK}[1]{\textcolor{red}{[OK]}}

\title[The impact of magnetic fields during TDEs]{The impact of magnetic fields during tidal disruption events}

\author[S. Pacuraru et al.]{
Simona Pacuraru,$^{1}$\thanks{E-mail: sxp1082@student.bham.ac.uk}
Clément Bonnerot$^{1}$ and Martin E. Pessah$^{2}$
\\
$^{1}$School of Physics and Astronomy \& Institute for Gravitational Wave Astronomy, University of Birmingham, Birmingham, B15 2TT, UK\\
$^{2}$ Niels Bohr International Academy, Niels Bohr Institute, Blegdamsvej 17, DK-2100 Copenhagen Ø, Denmark \\
}

\date{Accepted 2026 March 23. Received 2026 March 20; in original form 2025 November 26}

\pubyear{2015}

\begin{document}
\label{firstpage}
\pagerange{\pageref{firstpage}--\pageref{lastpage}}
\maketitle

\begin{abstract}
During a tidal disruption event (TDE) the stream debris inherits the magnetic field of the star. As the stream stretches, the magnetic field evolves and can eventually become dynamically important. We study this effect by means of magnetohydrodynamic simulations and a semi-analytic model of the disruption of a main-sequence star by a supermassive black hole. For stellar magnetic fields stronger than $\sim 10^4\,\rm{G}$, magnetic pressure becomes important in a significant fraction of the mass of the stream, leading to a fast increase in its thickness, an effect that may impact its subsequent evolution. We find that this dynamical effect is associated with a phase of transverse equilibrium between magnetic and tidal forces, which causes the stream width to increase with distance to the black hole as $H \propto R^{5/4}$. In the unbound tail, this fast expansion could enhance the radio emission produced by the interaction with the ambient medium, while in the returning stream, it may qualitatively affect the subsequent gas evolution, particularly the gas dynamics and radiative properties of shocks occurring after the stream's return to pericentre. By characterizing the magnetohydrodynamic properties of the stream from disruption to the first return to pericentre, this work provides physically motivated initial conditions for future studies of the later phases of TDEs, accounting for magnetic fields. This will ultimately shed light on the role of magnetic fields in enabling angular momentum transport in the ensuing accretion disk, thereby affecting observable signatures such as X-ray radiation and relativistic outflows. 
\end{abstract}




\section{Introduction}\label{S:Introduction}
Tidal disruption events (TDEs) occur when a star is torn apart by the strong gravitational pull of a supermassive black hole residing at the centre of a galaxy \citep[][]{Rees1988}. A full stellar disruption results in the formation of an elongated stream of gas, half of which becomes unbound, escaping the system at high velocities. The other half becomes bound and as it revolves around the black hole it undergoes relativistic apsidal precession which causes it to eventually self-intersect, initiating the formation of an accretion flow around the black hole \citep[e.g.][]{BonnerotStone2021}. 

During this event, a luminous flare of radiation is produced, observable at different wavelengths. The majority of TDE candidates have been discovered in the ultraviolet and optical bands \citep[e.g.][]{vanVelzen2021,Hammerstein2023}, although some also exhibit X-ray \citep[e.g.][]{Sazonov2021,Guolo_2024} and radio emission \citep[e.g.][]{Alexander2020,Goodwin2023}. In addition, a small subset of the observed TDEs shows evidence of relativistic jets \citep[e.g.][]{Bloom2011,Andreoni22}. To explain the very diverse observational features of these flares, TDEs have been extensively studied with both theoretical modelling and numerical simulations, which aim to provide a physical understanding of the origin of light during a TDE and, ultimately, use these transient events to probe the properties of massive black holes.

Simulating a TDE from stellar disruption to disk formation is computationally very challenging. The most uncertain stage of the evolution is the first return of the bound part of the stream to pericentre. Here the gas undergoes a strong tidal compression, which drives it to very small spatial scales \citep[][]{Kochanek1994}, making the problem computationally prohibitive. Because of this limitation, TDEs have been studied with both global hydrodynamic simulations \citep[e.g.][]{Ryu2023,Steinberg2024,Price_2024}, which can capture the overall dynamics from disruption to disk formation, and local simulations \citep[e.g.][]{Jiang2016, BonnerotLu2022, Huang_2024}, which aim to resolve the small scale dynamics by focusing on individual stages of the event. These include, the evolution of the stream around the black hole, the strong tidal compression near pericentre, the subsequent stream self-intersection driven by relativistic precession, and finally, the formation of the disk. 

In addition to the computational challenge, there are physical processes inherent to TDEs which are often neglected for simplicity. In this paper, we focus on the effect of magnetic fields. The magnetic field involved during a TDE originates from the star undergoing the tidal disruption, and its presence can have important consequences at different stages during the evolution. The first TDE simulations to address this are those by \citet{Guillochon2017} and \citet{Bonnerot2017}, which focused on the very early stages of the event, since the star's approach towards the black hole up to a few tens of hours after disruption. These works find that the stellar magnetic field gets deformed and stretched in response to the tidal field of the black hole, and it may later become dynamically important in the stream. This is expected to cause a fast increase in its width, potentially affecting the later evolution. \citet[][]{Meza25} find that, during the stream self-intersecting collisions, this thickness increase can reduce the acceleration of the outflowing gas, potentially affecting the circularization process. 

Regardless of their dynamical impact at early times, magnetic fields are thought to be essential for driving accretion when the disk is formed, as they enable angular momentum transport via the magnetorotational instability (MRI) \citep[][]{1991ApJ...376..214B}. Consequently, magnetic fields may be involved in powering the observed X-ray emission, expected to originate from accretion \citep[][]{1973A&A....24..337S,Mummery2020}. Despite these expectations, several TDE simulations that include magnetic fields during disk formation find that the MRI is subdominant compared to purely hydrodynamical effects \citep[e.g.][]{sadowski2018,Curd2023,Meza25}. A more recent simulation by \citet{abolmasov2025}, finds that the disk formed during the event is subject to significant MHD turbulence and magnetic field amplification. Magnetic fields are also invoked to explain the formation of relativistic jets through the Blandford-Znajek mechanism \citep[][]{BlandfordZnajek}, which requires the accumulation of magnetic flux at the black hole horizon. However, the magnetic flux required to account for the jet power inferred from TDEs is orders of magnitude larger than that of the original star (see e.g. \citet{KrolikPiran2012}, \citet{Kelley2014} and \citet{Bradnik2017} for possible solutions to this discrepancy). Investigating the role of magnetic fields in these later stages of the event requires a detailed characterization of the MHD properties of the stream at early times.

In this paper, we study for the first time the effect of the dynamical impact of magnetic fields during a TDE, from the time of disruption up until the first return of the stream close to pericentre. 
We employ MHD simulations using the code \textsc{gizmo} \citep[][]{Hopkins2015-gizmo} and the semi-analytic model by \citet{Bonnerot2022a}, which we extend to include the effect of magnetic fields. By characterizing the evolution of the MHD properties of the stream in the early stages of the event, this work provides physically motivated initial conditions for studying, with the inclusion of magnetic fields, the later phases of a TDE, e.g. the nozzle shock driven by the tidal compression near pericentre, the self-crossing shock and the resulting disk formation. We estimate the region of parameter space where we expect the magnetic field to become dynamically important during these early stages of the event. We quantify in detail its effect on the transverse expansion of the stream, finding deviations from the hydrodynamic evolution within $1 \, \rm{year}$ after disruption if the stellar magnetic field is above $\sim 10^{4} \, \rm{G}$. This stream expansion can have important consequences for the later evolution of the TDE, affecting, for instance, the hydrodynamic and radiative properties of shocks occurring after the stream's infall towards the black hole. 

The outline of the paper is as follows. In Section~\ref{S:Methods} we describe the semi-analytic model and the numerical setup used to perform the simulations. Section~\ref{S:The dynamical impact of magnetic fields} presents the results on the dynamical impact of the magnetic field and its effect on the stream thickness. In Section~\ref{S:MHD properties of TDE streams} we describe the results on the MHD properties of the stream. In Section~\ref{S:Discussion} we describe some caveats in our approach and the implications of this work for the later stages of a TDE. Finally, in Section~\ref{S:Conclusions} we summarize the paper and give conclusions.

\section{Methods}\label{S:Methods}
We consider the tidal disruption of a magnetized main-sequence star of mass $M_{\star}=\rm{M_{\odot}}$ and radius $R_{\star} = \rm{R_{\odot}}$ by a supermassive black hole of mass $M_{\rm{h}}=10^6\,\rm{M_{\odot}}$, where $M_{\odot}$ and $R_{\odot}$ are, respectively, the mass and radius of the sun. The star is modelled as a polytropic sphere of gas with polytropic index $\Gamma = 5/3$ and it is evolved assuming an adiabatic equation of state, $P\propto \rho^{\gamma}$, where $\gamma=5/3$ is the adiabatic index. Its centre of mass is initially set on a parabolic trajectory around the black hole. The pericentre radius of the orbit $R_{\rm{p}}$ is set equal to the tidal radius $R_{t} = R_{\star}(M_{\rm{h}}/M_{\star})^{1/3}$, that is the separation below which the tidal pull of the black hole exceeds the self-gravity within the star. 

We evolve the star assuming ideal magnetohydrodynamics (MHD), which greatly simplifies the physical interpretation of the results. This assumption holds as long as the ionization fraction is high enough to treat the gas as a perfect conductor. Initially, while the star is in hydrostatic equilibrium, the temperature can be approximated as $T_{\star}\approx GM_{\star}m_{\rm p}/(k_{B}R_{\star})=10^7\, \rm{K}$, where $G$ is the gravitational constant, $k_{B}$ is the Boltzmann constant and $m_{\rm p}$ is the proton mass. At this temperature the gas can be assumed fully ionized. However, the disruption causes the gas to adiabatically cool down as the debris stream expands. When the temperature reaches a value of $T\sim10^4\,\rm{K}$, hydrogen starts to recombine, which can cause the gas to deviate from the ideal MHD evolution. We will discuss the importance of these deviations and the timescales over which they become relevant in Section~\ref{S:Discussion}.

We employ both global numerical simulations and a semi-analytic model. These two methods turn out to be complementary tools for a full comprehension of this problem. The numerical simulation treats the magnetohydrodynamics very accurately, but it is computationally limiting. The semi-analytic model is less accurate, but allows to explore more of the parameter space with lower computational expense. In the following we describe the details of these two methods.

\subsection{Magnetohydrodynamic Simulations}
\subsubsection{Numerical setup}
The tidal disruption simulations are performed using the Meshless Finite Mass version of the code \textsc{gizmo} developed by \citet{Hopkins2015-gizmo}. The gravity implementation is based on the N-body solver of \textsc{gadget-3} \citep[][]{Springel2005} and it includes an adaptive gravitational softening. To evolve the magnetic field we solve the fluid equations using the ideal MHD module developed by \citet{Hopkins2015}. To minimize the divergence of the magnetic field we use the intermediate version of the constrained gradient method developed by \citet{Hopkins2016} which adds some terms to the usual divergence cleaning methods \citep[][]{DEDNER2002645,Powell1999}, to further reduce divergence errors.

The initial star is numerically modelled as a three-dimensional distribution of gas particles that resembles a polytropic sphere in hydrostatic equilibrium. We generate this initial condition using the routine developed as part of the code \textsc{phantom} \citep[][]{Price2018-phantom}, which samples the coordinates of the particles so that the density profile is the solution of the Lane-Emden equation for a polytrope with $\Gamma=5/3$. We run one hydrodynamic (HD) and several MHD simulations, where we vary the geometry and the strength of the magnetic field. We set the resolution to $10^{5}$ particles. We ran a higher resolution simulation with $10^6$ particles, and did not notice any significant difference in the results.

We first evolve the star in isolation using \textsc{gizmo} to relax the initial condition to an equilibrium configuration. We then place the star at a distance of $3\,R_{\rm{t}}$ from the black hole, where the tidal pull of the black hole is negligible compared to the stellar self-gravity, and initialize its velocity such that its centre of mass moves towards the black hole on a parabolic orbit. We follow the evolution of the bound part of the stream up until it comes back to pericentre. The study of the subsequent evolution is out of the scope of this work, so we eliminate from the domain particles that fall below a distance of $5\,R_{\rm{t}}$, where the resolution becomes so poor that the evolution is significantly affected by numerical errors. 

\subsubsection{Initial Magnetic Field}

The internal magnetism of main-sequence stars has not been accurately measured. Our knowledge is restricted to the stellar surface, where typical magnetic fields are in between $\sim 10\,\rm{G}$ in the case of sun-like stars \citep[e.g.][]{1947ApJ...105..105B}, and $\sim 10\,\rm{kG}$ in the case of chemically peculiar stars \citep[][]{Mathys1997}. These surface strengths are expected to be a factor of $\sim 10$ smaller than the interior magnetic field \citep[][]{2023Fuller}. This is furthermore consistent with recent asteroseismology observations of the internal magnetic fields in red giants \citep[][]{2022Natur.610...43L,2023A&A...670L..16D,Hatt2024}, which suggest minimal strengths in the range $\sim 1\,\rm{kG}-10\,\rm{kG}$ during the main-sequence. 

Given the uncertainty in the initial magnetic field strengths of stars, in this work we explore a range of initial magnetic field values. In our numerical simulations we consider the following cases: $B_{\star}=1\,\rm{G},0.1\,\rm{MG},1\,\rm{MG},3\,\rm{MG},10\,\rm{MG}$. To explore the rest of the parameter space we make use of the semi-analytic model. The choice to include strongly magnetized stars is motivated both by computational convenience and by physical constraints on the initial magnetic field required to have any effect on the evolution. We find that during a TDE the magnetic field becomes dynamically important earlier for more strongly magnetized stars. As a result, simulating its dynamical effects on the stream is computationally less demanding if the star is initially highly magnetized. 

\begin{figure}
    \includegraphics[width=1\columnwidth]{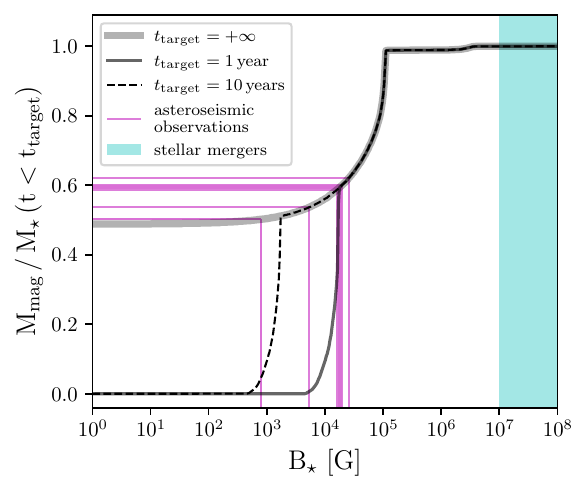}
	\caption{Fraction of the mass in the stream that becomes magnetically supported within time $t_{\rm target}$ after disruption, as a function of the initial magnetic field. The grey and black lines are obtained as described in Section~\ref{ss:Onset of magnetic pressure support}. The pink lines correspond to the magnetic fields of main sequence stars inferred from asteroseismology measurements of red giants (see \citet{2023A&A...670L..16D}), while the blue shaded region represents the magnetic field found in simulations of stellar mergers (\citet{Ryu_2025}, \citet{Scheider2019}).} 
    \label{fig:mass_fraction}
\end{figure}

In Section~\ref{ss:Onset of magnetic pressure support}, we analytically estimate the stream's mass where the magnetic field becomes dynamically important after a given target time as a function of the stellar magnetic field. The result is illustrated in Fig.~\ref{fig:mass_fraction}, which shows (see grey solid line) that the magnetic field in the star must be close to $10^4 \, \rm{G}$ for it to become dynamically important in at least half of the mass of the stream within a year after disruption. Although it is uncertain to date whether main sequence stars possess magnetic fields this strong, asteroseismic observations of red giants suggest that such strengths can exist during the main sequence (pink lines, obtained from \citet{2023A&A...670L..16D}). Furthermore, galactic nuclei offer a favourable environment for amplifying the stellar magnetic field through stellar mergers \citep[][]{Bradnik2017}. Numerical simulations have shown that these mergers can enhance the magnetic field by many orders of magnitude, reaching $\sim 10^7-10^8\,\rm G$ \citep[][]{Scheider2019,Ryu_2025}, in which case we expect the entire stream to be dynamically affected by magnetic fields (blue shaded region).

The topology of stellar magnetic fields is also uncertain. Currently, a main theoretical constraint comes from the requirement that a stellar magnetic field has to be in stable equilibrium during the star's lifetime. Numerical simulations by \citet{2004Natur.431..819B} have shown that such stability can be achieved by mixed poloidal and toroidal fields. Instead, purely toroidal and purely poloidal magnetic fields are unstable and decay rapidly. To study the dependence of our results on the magnetic field geometry, we consider three distinct configurations: (i) a simple uniform field, (ii) a toroidal field in the plane of the orbit, and (iii) a more realistic configuration that mimics the tangled field geometry found by \citet{2004Natur.431..819B}. For the latter we use the analytic equilibrium solution by \cite{Kamchatnov1982}
\begin{equation}\label{eq:B_Kamchatnov}
    B_{\rm{x_{\star}}}=\frac{2(x_{\star}z_{\star}-y_{\star})}{(1+r^{2})^{3}}; \hspace{0.2cm}B_{\rm{y_{\star}}}=\frac{2(y_{\star}z_{\star}+x_{\star})}{(1+r^{2})^{3}}; \hspace{0.2cm}B_{\rm{z_{\star}}}=\frac{1-x_{\star}^{2}-y_{\star}^{2}}{(1+r^{2})^{3}},
\end{equation}
where $x_{\star}$, $y_{\star}$ and $z_{\star}$ are cartesian coordinates and $r=\sqrt{x_{\star}^{2}+y_{\star}^{2}+z_{\star}^{2}}$ is the distance from the centre of the star. We start by setting the $x_{\star}$- and $y_{\star}$-axis in the plane of the orbit, with the $y_{\star}$-axis parallel to the initial velocity of the center of mass of the star. In a realistic scenario, the stellar magnetic field will be randomly oriented with respect to the orbital plane. With the purpose of having a more general magnetic field that accounts for this, we rigidly rotate the field defined in equation~(\ref{eq:B_Kamchatnov}) about the $x_{\star}$-axis, so that the $y_{\star}$-axis is inclined by $45^{\circ}$ with respect to the orbital plane.

Since the initial field is defined only at the particles’ locations and vanishes outside the stellar surface, neither the uniform nor the tangled magnetic fields are divergence-free at the edges of the star. As a result, in a small fraction of the gas, corresponding to roughly $5 \%$ of the total mass, the divergence error $h |\nabla \cdot \textbf{B}|/|\textbf{B}|$ is of order unity, where $h$ is the smoothing length. Importantly, the evolution of this material does not appear to be significantly impacted, as will be shown in the results through comparison with simple trends expected from magnetic flux conservation. Away from these boundaries the average divergence error remains smaller than $\approx 10 \%$ at all times.

\subsection{Semi-analytic model of a magnetized TDE}
\label{sec:Semi-analytic model of a magnetized TDE}
We employ the semi-analytic model developed by \citet{Bonnerot2022a}, which we extend to include magnetic fields. The main scope of the model is to draw conclusions on the dynamical impact of the magnetic field for the part of the parameter space that is not accessible with full numerical simulations.
 
The star is assumed to travel on a parabolic orbit and its initial location is set equal to the tidal radius, based on the frozen-in approximation. To capture the diverse hydrodynamic evolution of different stream elements, the star is sliced into two-dimensional sections located at different distances $r=\mu R_{\star}$ from the centre of mass of the star, where $-1\leq\mu\leq1$ is a "boundness" parameter describing how bound the trajectory of a section is, with $\mu<0$ corresponding to bound gas and $\mu>0$ to unbound gas. In particular one can express the orbital energy of these sections as $\epsilon=\mu \Delta \epsilon$, where $\Delta \epsilon = GM_{\rm h}R_{\star}/R^{2}_{\rm{t}}$ is the spread in orbital energy imparted to the debris. As depicted in figure 1 of \citet{Bonnerot2022a}, the geometry of the problem is described in terms of a basis of unit vectors ($\textbf{e}_{\parallel}$, $\textbf{e}_{\perp}$, $\textbf{e}_{\rm{z}}$) that is co-moving with the centre of mass of a given section, where $\textbf{e}_{\parallel}$ and $\textbf{e}_{\perp}$ are aligned with the orbital plane, while $\textbf{e}_{\rm{z}}$ is orthogonal to it. The hydrodynamic evolution of a section with a given boundness $\mu$ is described by a set of ordinary differential equations (see equations (3), (4), (5), (6), (9) and (13) in \citet{Bonnerot2022a}) that approximate the dynamics. The sections are assumed to be elliptical and featured by a vertical width $H$, an in-plane width $\Delta$ and a local elongation $\ell$. The mass density of a section is defined as $\rho =\Lambda/(\pi H \Delta)$, where $\Lambda=m/\ell$ is the linear density and $m$ is the mass of the section.

In this work we generalize the model to include the effect of the magnetic force $\textbf{f}_{\rm{mag}}=-\nabla B^{2}/(8\pi)+(\textbf{B}\cdot \nabla)\textbf{B}/(4\pi)$ on the stream. The magnetic field is decomposed as $\textbf{B}=B_{\parallel}\textbf{e}_{\parallel}+B_{\perp}\textbf{e}_{\perp}+B_{z}\textbf{e}_{z}$. With the inclusion of this magnetic field the equations of motion for the vertical and horizontal widths are given by
\begin{equation}
\begin{split}
    &\ddot{H} = \ddot{H}_{\rm{g}} + \ddot{H}_{\rm{p}} + \ddot{H}_{\rm{t}} + \ddot{H}_{\rm{m}}\\
    &\ddot{\Delta} = \ddot{\Delta}_{\rm{g}} + \ddot{\Delta}_{\rm{p}} + \ddot{\Delta}_{\rm{t}} + \ddot{\Delta}_{\rm{m}}.
    \label{eq:ddot_H}
\end{split}
\end{equation}
In both equations, the first three terms are, from left to right, the self-gravity force, the pressure force and the tidal force, and their evolution is described by equations (7), (8), (10), (11) and (12) in \citet{Bonnerot2022a}. In this work we introduce the last term, that corresponds to the transverse magnetic acceleration (see derivation in Appendix~\ref{app:Magnetic acceleration equations})
\begin{equation}
        \ddot{H}_{\rm{m}} = \frac{1}{\rho}(\textbf{f}_{\rm{mag}} \cdot \textbf{e}_{z}) = \frac{1}{4\pi\rho}\left(\frac{B^{2}}{2H} - \frac{B^{2}_{\rm{z}}}{2H}- \frac{B_{\perp}B_{\rm{z}}}{\Delta}\right),
        \label{eq:ddot_Hm}
\end{equation}
\begin{equation}
        \ddot{\Delta}_{\rm{m}} = \frac{1}{\rho}(\textbf{f}_{\rm{mag}} \cdot \textbf{e}_{\perp}) = \frac{1}{4\pi\rho}\left(\frac{B^{2}}{2\Delta} - \frac{B^{2}_{\perp}}{ 2\Delta} - \frac{B_{\perp}B_{\rm{z}}}{H }\right)
        \label{eq:ddot_Dm},
\end{equation} 
where $B^2 = B^2_{\parallel}+B^2_{\perp}+ B^2_{z}$.

Because of the ideal MHD assumption, each component of the field is inversely proportional to the surface area orthogonal to its direction, such that $B_{\parallel}$, $B_{\perp}$ and $B_{z}$ evolve proportionally to $1/H\Delta$, $\Lambda/H$ and $\Lambda/\Delta$, respectively.  This leads to the following equations for the evolution of the magnetic field
\begin{equation}
        B_{\parallel} = B_{\parallel,\rm{i}} \frac{H_{\rm{i}}\Delta_{\rm{i}}}{H\Delta};\hspace{0.2cm} B_{z} = B_{z,\rm{i}} \frac{\Lambda\Delta_{\rm{i}}}{\Lambda_{\rm{i}}\Delta};\hspace{0.2cm} B_{\perp} = B_{\perp,\rm{i}} \frac{\Lambda H_{\rm{i}}}{\Lambda_{\rm{i}} H},
        \label{eq:B}
    \end{equation}
where $H_{\rm{i}}=\Delta_{\rm{i}}=R_{\star}(1-\mu^{2})^{1/2}$ are the initial vertical and horizontal widths, $\Lambda_{\rm i}$ is the initial linear density and $B_{\parallel,\rm{i}}, B_{z,\rm{i}}, B_{\perp,\rm{i}}$ are the initial components of the magnetic field. 

One important assumption of the model is that it is "one-zone", meaning that we only evolve one average quantity (e.g. density, field strength) for each section, while realistically there should be a profile. Furthermore, the sections are defined to be orthogonal to the longitudinal direction. Because of this, fluid elements away from the centre of mass continuously cross the section due to shearing. However, this effect does not affect the gas properties, due to their local invariance along the stream's longitudinal direction. In particular, the magnetic field is assumed to be locally homogeneous enough such that particles crossing the section retain the same magnetic field as those that left. Despite this approximation, we show below that the semi-analytical model is able to recover the qualitative behavior found in our simulations. Finally, we note that, because of magnetic flux conservation, the divergence of the magnetic field is conserved, and therefore it suffices to ensure that it is zero initially.

\section{Results}\label{S:results}

In this section, we present the results on the evolution of the MHD properties of the debris stream produced during a TDE, with particular emphasis on how magnetic fields can drive deviations from the hydrodynamic behavior. To facilitate the comparison, we first provide a brief summary of the current understanding of the evolution in the absence of magnetic fields.

\subsection{Overview of the hydrodynamic evolution}\label{S:Overview of the hydrodynamic evolution}
The stream can be understood as a series of cylinders with elongation $\ell$ and transverse sizes $H$ and $\Delta$ \citep[][]{Bonnerot2022a}, and whose center of mass trajectory is characterized by the boundness parameter $\mu\equiv \epsilon/\Delta\epsilon$\footnote{We note that the orbital energy distribution in the numerical simulation is wider than in the semi-analytic
model (see e.g. \citet{Lodato2009}), which results in the boundness reaching values $|\mu|>1$.}. Typically $H\approx\Delta$ at all times we consider, so we only comment on $H$ hereafter. While the evolution of the stream's length $\ell$ is entirely ballistic, self-gravity and pressure forces may also play a role in the evolution of the transverse width $H$ \citep[][]{Kochanek1994,Coughlin2015}. One can distinguish two different regimes for the transverse evolution, which are separated by a critical density $\rho_{\rm crit}\approx M_{\rm h}/(2\pi R^{3})$, corresponding to a critical boundness $\mu\equiv\mu_{\rm{crit}}$\footnote{The resulting value of $\mu_{\rm crit}$ is different between model ($\mu_{\rm crit}\approx0.533$) and simulations ($\mu_{\rm crit}\approx1$), which is expected because the density is computed differently in the two methods.} \citep[][]{Coughlin2016,Bonnerot2022a}: sections with $|\mu|<\mu_{\rm{crit}}$ (hydrostatic regime) are initially sufficiently dense to remain in hydrostatic equilibrium in the transverse direction soon after disruption, while for $|\mu|>\mu_{\rm{crit}}$ (ballistic regime), the stream elements are so dilute that the tidal pull dominates self-gravity. 

Shortly after the disruption, the length of a stream section evolves with the distance from the black hole $R$ following
\begin{equation}\label{eq:l_R2}
    \ell \propto R^{2}
\end{equation}
everywhere in the stream (\citet{Coughlin2016}), while the width scales as
\begin{equation}\label{eq:H_hydro_early}
    H, \Delta \propto \begin{cases} R^{1/2} & \mbox{hydrostatic regime} \\ R & \mbox{ballistic regime}. 
\end{cases}
\end{equation}
Later, when the bound part of the stream starts to return towards pericentre, the gas becomes so dilute that the tidal force dominates the dynamics. In these conditions, the evolution of the elongation transitions to
\begin{equation}\label{eq:ell_propto_R-12}
    \ell \propto R^{-1/2},
\end{equation}
while the width keeps following equation~(\ref{eq:H_hydro_early}). However, here the scaling $H \propto R^{1/2}$ is not due to hydrostatic balance, but ballistic motion (see \citet{Bonnerot2022a}).  

In the unbound part of the stream, equations~(\ref{eq:l_R2}) and~(\ref{eq:H_hydro_early}) apply before the gas is so far away that it reaches its terminal velocity. From that point on, the elongation follows
\begin{equation}\label{ell_propto_R}
    \ell \propto R,
\end{equation}
and the width evolves as
\begin{equation}\label{eq:H_hydro_late}
    H,\Delta \propto \begin{cases} R^{1/4} & \mbox{hydrostatic regime} \\ R & \mbox{ballistic regime}.
\end{cases}
\end{equation}

\subsection{The dynamical impact of magnetic fields}\label{S:The dynamical impact of magnetic fields}
In this section we describe how magnetic fields can become dynamically important during the early stages of a TDE, thereby affecting the evolution of the width of the stream debris. In Table~\ref{tb:scalings} we summarize the evolution of the width in different regimes and in different parts of the stream, by showing the scalings known from previous works for the pure hydrodynamic evolution, and the ones found in this work when magnetic fields dominate.

\begin{table}
\caption{Approximate scalings for the elongation and the width in different portions of the stream, in the hydrostatic (hyd) and in the ballistic regime (bal). $H_{\rm hd}$ is the width in the absence of magnetic fields, known from previous works (see Section~\ref{S:Overview of the hydrodynamic evolution}). $H_{\rm mhd}$ is the width in the case where the magnetic field becomes dynamically important and the corresponding scalings were obtained in this work and are described in Section~\ref{Ss:Stream width increase under magnetic pressure}.}

\begin{threeparttable}
\centering
\resizebox{\linewidth}{!}{
\begin{tabular}{l@{\hspace{5pt}}c|c|c c|c c}
\toprule
 &  & $\ell$
 & \multicolumn{2}{c|}{$H_{\rm hd}$}
 & \multicolumn{2}{c}{$H_{\rm mhd}$} \\
\midrule

$\mu=0$ 
  &  
  & $R^{2}$ 
  & \multicolumn{2}{c|}{$R^{1/2}$\tnote{**}}
  & \multicolumn{2}{c}{$R^{5/4}$} \\
\midrule

\multirow{4}{*}{$\mu>0$}
  & \hspace{-0.85cm}\textbf{Regime}
  & 
  & \textbf{hyd}
  & \textbf{bal}
  & \textbf{hyd}
  & \textbf{bal} \\[2pt]

\midrule
  & $t \ll t_{\rm tr}$ 
  & $R^{2}$
  & $R^{1/2}$ & $R$
  & $R^{5/4}$ & $R^{5/4}$ \\

  & $t \gg t_{\rm tr}$ 
  & $\hspace{-0.15cm}R$
  & $R^{1/4}$ & $R$
  & $\hspace{-0.35cm}R$ & $\hspace{-0.35cm}R$ \\
\midrule

\multirow{2}{*}{$\mu<0$}
  & $t \ll t_{\rm apo}$ 
  & $R^{2}$
  & $R^{1/2}$ & $R$
  & $R^{5/4}$ & $R^{5/4}$ \\

  & $t \gg t_{\rm apo}$ 
  & $R^{-1/2}$
  & $R^{1/2}$\tnote{*} & $R$
  & $R^{1/2}$\tnote{*} & $R$$\text{ or }R^{5/8}$ \\
\bottomrule

\end{tabular}
}
    \begin{tablenotes}
    \item[*] In the infalling part of the stream the scaling for the width is less certain. This is because, as a result of the tidal force, the width evolution can vary between $H \propto R^{1/2}$ and $H \propto R$ depending on the orientation of the line where the orbital planes of the stream elements cross (see \citet{Bonnerot2022a}).
    \item[**] This scaling is valid as long as the parabolic stream element is maintained in hydrostatic equilibrium. It is uncertain whether this remains valid indefinitely. 
    \end{tablenotes}
    \end{threeparttable}
\label{tb:scalings}
\end{table}

\subsubsection{Magnetic field alignment with stretching direction}\label{sec:Magnetic field re-alignment}
In Fig.~\ref{fig:splash_and_Bstream} we show the evolution of the magnetic field strength and geometry in the simulation with an initial uniform field $B_{\star}=3\,\rm{MG}$ aligned with the centre of mass velocity at the initial time when the star is located at $R=3\,R_{\rm t}$. The magnetic field strength is not substantially modified before the pericentre passage, reflecting the fact that the star is not subject to significant compression or expansion during the infall phase. The stream starts to get significantly stretched upon reaching the tidal radius (at $t\approx1.5\,\rm{h}$ in our simulation, corresponding to the top right panel of Fig.~\ref{fig:splash_and_Bstream}), which is associated with an alignment of the magnetic field lines in the direction of stretching, as previously found by \citet{Guillochon2017} and \citet{Bonnerot2017}. This alignment of the magnetic field is visible in our simulation within a few hours after the first pericentre passage (see bottom panel of Fig.~\ref{fig:splash_and_Bstream}) and it can be physically understood by evaluating the relative importance between the components of the field in the longitudinal and transverse directions. Because of flux conservation (equation~(\ref{eq:B})), the ratio between the horizontal and the longitudinal components of the field follows $B_{\perp}/B_{\parallel}\propto\Delta/\ell$, using $\ell=m/\Lambda$ ($\ell$ will be defined like this hereafter). At early times the elongation evolves as $\ell \propto R^{2}$ (equation~(\ref{eq:l_R2})), while the horizontal width follows $\Delta\propto R^{1/2}$ and $\Delta\propto R$ in the hydrostatic and in the ballistic regime, respectively (equation~(\ref{eq:H_hydro_early}) with $\Delta\approx H$). Using these scalings we obtain
\begin{equation}\label{B_paralel_perp_ratio}
    \frac{B_{\perp}}{B_{\parallel}} \propto \begin{cases} R^{-3/2} & \mbox{hydrostatic regime} \\ R^{-1} & \mbox{ballistic regime},
\end{cases}
\end{equation}
which shows that $B_{\perp}$ decreases over time much faster than $B_{\parallel}$ everywhere in the stream. Since $H \approx \Delta$, $B_{z}$ follows the same scaling as $B_{\perp}$, again as a result of magnetic flux conservation. This reduction in both transverse magnetic field components implies that the field will tend to align with the direction of stretching, provided that $B_{\parallel}$ is initially non-zero.
\subsubsection{Onset of magnetic pressure support}\label{ss:Onset of magnetic pressure support}

\begin{figure}
    \includegraphics[width=1\columnwidth]{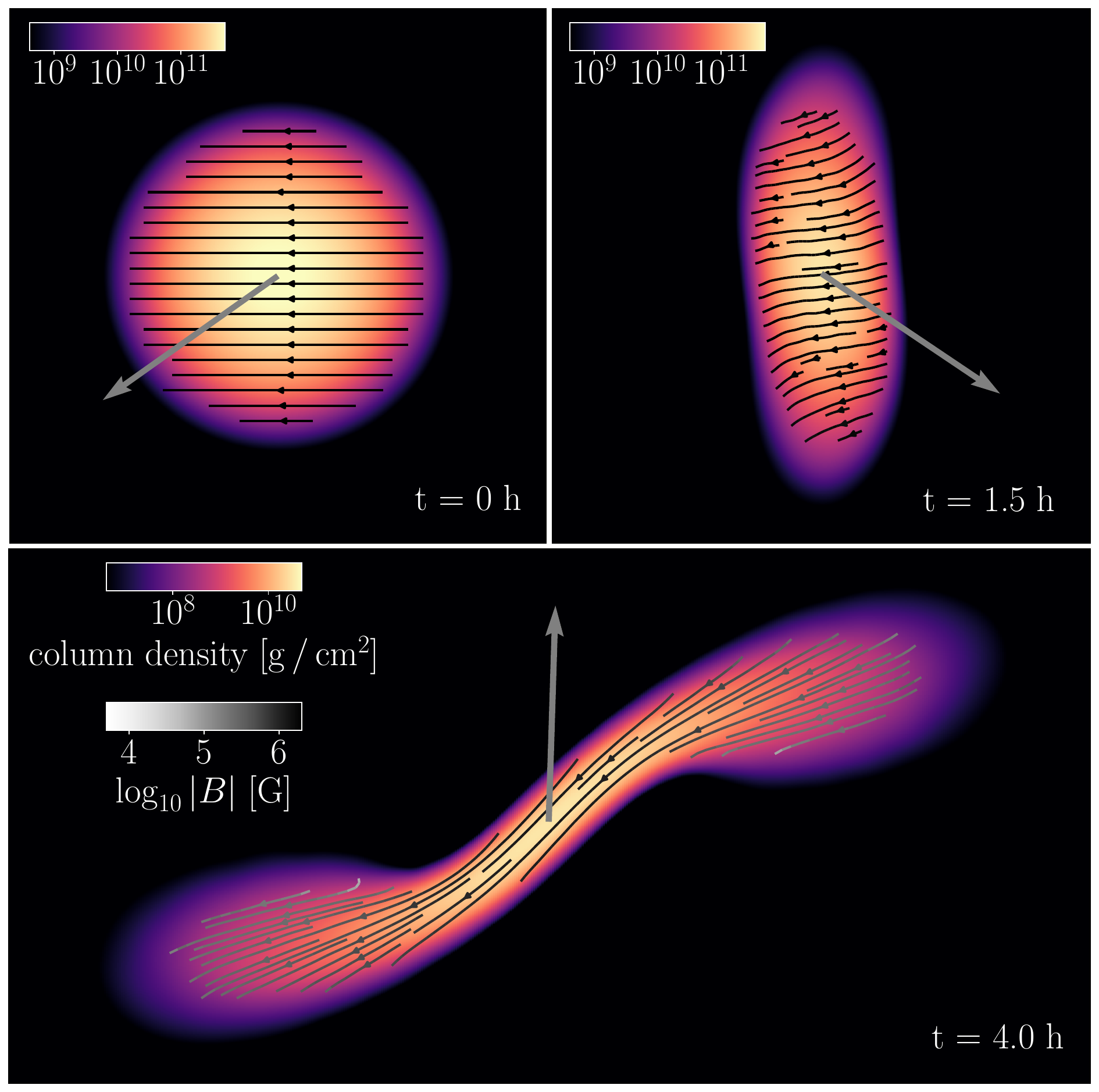}
	\caption{Snapshots showing the projection of the density and magnetic field on the orbital plane in the simulation initialized with a uniform field $B_{\star}=3\,\rm{MG}$ at times $t=0\, \rm{h}$, $t=1.5\, \rm{h}$ and $t=4\, \rm{h}$, corresponding respectively to distances from the black hole of $R=3\,R_{\rm t}$, $R=R_{\rm t}$ and $R=4.5\,R_{\rm t}$. The gray arrows point towards the black hole. As the star gets deformed and stretched the magnetic field aligns with the elongation direction. The colorbar describing the magnetic field strength is identical in the three panels.} 
    \label{fig:splash_and_Bstream}
\end{figure}

\begin{figure*}
    \includegraphics[width=1\textwidth]{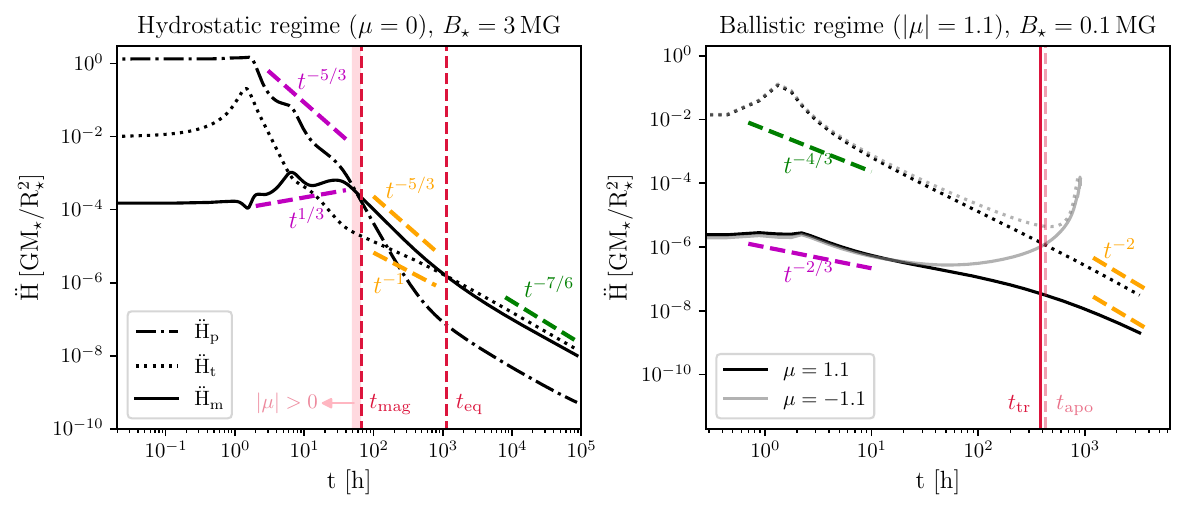}
    \caption{Left panel: evolution of accelerations $\ddot{H}_{\rm{m}}$, $\ddot{H}_{\rm{t}}$ and $\ddot{H}_{\rm{p}}$, for a parabolic stream element in the simulation initialized with a uniform magnetic field $B_{\star}=3\,\rm{MG}$. The purple dashed segments indicate the expected scalings for $\ddot{H}_{\rm{m}}$ and $\ddot{H}_{\rm{p}}$  during the early evolution, when the stream is still confined by self-gravity. The leftmost red vertical line marks the time $t_{\rm{mag}}$ at which  $\ddot{H}_{\rm{m}}=\ddot{H}_{\rm{p}}$ in the simulation. After $t_{\rm{mag}}$ the stream element expands at constant speed causing $\ddot{H}_{\rm{m}}$ and $\ddot{H}_{\rm{t}}$ to follow the orange dashed segments. Later on, the tidal and magnetic accelerations reach an equilibrium with $\ddot{H}_{\rm{t}}=\ddot{H}_{\rm{m}}$ (rightmost vertical red line), after which they both follow the green dashed segments. Right panel: evolution of $\ddot{H}_{\rm{m}}$ and $\ddot{H}_{\rm{t}}$ for stream elements of boundness $\mu=-1.1$ and $\mu=1.1$ in the simulation initialized with a uniform magnetic field $B_{\star}=10^5\,\rm{G}$. The vertical red solid line marks the transition time $t_{\rm{tr}}$ (equation~(\ref{eq:t_tr})), while the vertical red dashed line indicates the time to reach apocenter $t_{\rm{apo}}$.}
    \label{fig:Pg_Pm_time}
\end{figure*}

To investigate the potential effects of the magnetic field during the early evolution of a TDE, in this section we track the evolution of the magnetic pressure force and we compare it with the other forces acting on the stream. We start by illustrating in the left panel of Fig.~\ref{fig:Pg_Pm_time} the temporal evolution of the transverse accelerations associated with magnetic pressure (solid line), tidal force (dotted line), and gas pressure (dash-dotted line), computed, respectively, as $\ddot{H}_{\rm{m}}=B^2/(\rho H)$, $\ddot{H}_{\rm{t}}=GM_{\rm h}H/R^3$ and $\ddot{H}_{\rm{p}}=P/(\rho H)$ in the simulation initialized with a uniform magnetic field $B_{\star}=3\,\rm{MG}$ for a parabolic stream section with $\mu=0$. To compute these different terms in the simulation, we calculate $\mu$ for each particle and slice the stream into thin sections of gas selected from a narrow range of orbital energies $\mu \pm \Delta\mu$. The density and the pressure are then averaged over the selected particles. To estimate the width we compute the distance of each particle contained in the section with respect to the centre of mass and we project and average these distances along the vertical direction to obtain $H$. 

The different components of the acceleration follow scalings in time (dashed segments in Fig.~\ref{fig:Pg_Pm_time}) that can be understood analytically as follows. The self-gravity and the pressure accelerations initially dominate the dynamics of the parabolic stream element, causing its width to increase as $H \propto R^{1/2}$ such that hydrostatic equilibrium is maintained (see equation~(\ref{eq:H_hydro_early})). As a result, the gas pressure and the self-gravity accelerations decrease with time according to the following scaling
\begin{equation}
    \ddot{H}_{\rm{g}} \propto \ddot{H}_{\rm{p}} \propto \frac{P}{\rho H} \propto R^{-5/2}\propto t^{-5/3},
    \label{dotHp}
\end{equation}
where we have used that $P \propto \rho^{5/3}$ and $\rho \propto (H \Delta \ell)^{-1} \propto R^{-3}$. The last proportionality in equation~(\ref{dotHp}) uses the scaling relation $R\propto t^{2/3}$ valid for a near-parabolic orbit \citep[][]{Sari_2010}. Shortly after pericentre passage, the magnetic field has aligned with the elongation direction (see Fig.~\ref{fig:splash_and_Bstream}), such that $B^2 \approx B^2_{\parallel}$, and the magnetic acceleration follows
\begin{equation}
    \ddot{H}_{\rm{m}} \approx \frac{B^{2} _{\parallel}}{\rho H} \propto \frac{l}{H^{2}\Delta} \propto R^{1/2} \propto t^{1/3},
    \label{dotHm}
\end{equation}
which shows that the magnetic force is expected to increase with time\footnote{We emphasize that the component of the field parallel to the stretching direction is crucial to the onset of this effect, as only this contributes to the increase in acceleration described by equation~(\ref{dotHm}). In contrast, the perpendicular component contributes a magnetic acceleration term that drops as $t^{-5/3}$, that is the same decay rate of gas pressure acceleration.}. These estimates are in good agreement with the result of our simulation, which shows that the gas pressure and the magnetic accelerations oscillate\footnote{These fluctuations are related to physical oscillations of the transverse size of the stream, that are excited by the tidal force during the first passage of the star at pericentre.} around the predicted scalings, such that $\ddot{H}_{\rm{m}}=\ddot{H}_{\rm{p}}$ at finite time after disruption, at which point the magnetic field becomes dynamically important, potentially driving deviations from the purely hydrodynamical evolution. 

In the following, we study in more detail the onset of this effect along the stream by evaluating the time $t_{\rm mag}$ over which the magnetic field becomes dynamically important as a function of $\mu$. This will allow us to estimate the region of parameter space in which the magnetic pressure support is expected to contribute significantly to the dynamics within timescales comparable to those of observed TDEs. 

\paragraph*{Hydrostatic regime ($|\mu|<\mu_{\rm crit}$).}
In this regime equations~(\ref{dotHp}) and~(\ref{dotHm}) still apply since the transverse widths and elongation follow the same scalings as in the parabolic case at early times. As a result, in this part of the stream, that accounts for $\sim 80 \%$ of the total mass, the plasma beta 
\begin{equation}\label{eq:betaM_def}
    \beta_{\rm{M}} \equiv \ddot{H}_{\rm{p}}/\ddot{H}_{\rm{m}},
\end{equation}
evolves as $\beta_{\rm{M}} \propto t^{-2}$, which leads to the following estimate for the time $t^{\rm{hyd}}_{\rm{mag}}$ after which the magnetic field is expected to become dynamically important in the hydrostatic regime:
\begin{align}\label{eq:t_mag}
    t^{\rm{hyd}}_{\rm{mag}} (\rm early)&= t_{\rm{str}}\beta_{\rm{M},\star}^{1/2} \notag\\
    &\approx 44\,\rm{h} \left(\frac{\beta_{\rm{M},\star}}{10^{4}}\right)^{1/2} \left(\frac{R_{\star}}{R_{\odot}}\right)^{3/2} \left(\frac{M_{\star}}{M_{\odot}}\right)^{-1/2},
\end{align}
where $\beta_{\rm{M,\star}}$ is the initial plasma beta and $t_{\rm{str}}\approx \sqrt{R^{3}_{\star}/GM_{\star}}$ is the time after which the stream has stretched significantly.\footnote{This is similar to the time spent at pericentre, which is also equal to the stellar dynamical time $\sqrt{R^{3}_{\star}/GM_{\star}}$ (for a pericentre equal to the tidal radius).} By this time the stream has reached a distance $R^{\rm hyd}_{\rm mag}=R_{\rm t}\beta^{1/3}_{\rm M,\star}\approx 22 \, R_{\rm t}(\beta_{\rm M,\star}/10^4)^{1/3}$ from the black hole. This estimate for $t^{\rm{hyd}}_{\rm{mag}}$ is consistent with the time after which $\ddot{H}_{\rm{p}}$ and $\ddot{H}_{\rm{m}}$ are equal in the parabolic section of our numerical simulation (see leftmost red vertical line in the left panel of Fig.~\ref{fig:Pg_Pm_time}). Equation~(\ref{eq:t_mag}) shows that the timescale for the onset of magnetic pressure support is independent of the black hole mass, and it is only a function of the magnetic properties of the star. Since  $t^{\rm{hyd}}_{\rm{mag}}$ increases with the plasma beta $\beta_{\rm{M},\star}$, the magnetic field is expected to become dynamically important earlier for more strongly magnetized stars. Moreover, $t^{\rm{hyd}}_{\rm{mag}}$ may vary with $\mu$ due to the fact that the plasma beta is generally not uniform across the star. In the simulation discussed in this section the magnetic field is initially homogeneous. Because gas pressure drops with $|\mu|$ (moving away from the stellar core), this implies that the plasma beta $\beta_{\rm{M},\star}$, and thus $t^{\rm{hyd}}_{\rm{mag}}$, are expected to decrease with $|\mu|$. This is consistent with the red shaded region drawn in Fig.~\ref{fig:Pg_Pm_time}, which shows the time at which $\ddot{H}_{\rm{m}}$ and $\ddot{H}_{\rm{p}}$ become equal in our simulation for the stream sections in the hydrostatic regime. We also note that the initial direction of the magnetic field has a negligible impact on the onset of magnetic pressure support, typically delaying it by a factor of a few.\footnote{Equation~(\ref{eq:t_mag}) was derived assuming the magnetic field to be initially aligned with the stretching direction. For an arbitrary magnetic field geometry, $B_{\parallel}=\cos(\theta) \,B$, where $\theta$ is the inclination angle between the magnetic field and the stretching direction, such that the plasma beta $\beta^{\parallel}_{\star}$ due to the parallel component of the magnetic field is $\beta^{\parallel}_{\star}=\beta_{\rm{M},\star}/\cos^{2}(\theta)$. As a result, the timescale $t_{\rm{mag}}$ estimated in equation~(\ref{eq:t_mag}) increases by a factor $\cos^{-1}(\theta)$. This means that, as long as the initial magnetic field is inclined with respect to the stretching direction by an angle $\theta < 70^{\circ}$, $t_{\rm mag}$ increases by a factor less than $\approx 3$.}

The scalings used to derive equation~(\ref{eq:t_mag}) apply at early times, but the stream's geometry evolves differently later on for non-parabolic stream elements, implying that the condition for the magnetic field to become dynamically important will change. In the case of bound stream elements that are initially in the hydrostatic regime ($-\mu_{\rm{crit}}<\mu<0$) this condition is modified at the apocentre $R_{\rm{apo}}\approx 2 \, a=R_{\rm{t}}|\mu|^{-1}(M_{\rm{h}}/M_{\star})^{1/3}$, where $a$ is the semi-major axis. As the elements start infalling towards the black hole, the scaling for the elongation changes to $\ell \propto R^{-1/2}$ (equation~(\ref{eq:ell_propto_R-12})). As a result, the ratio between magnetic force and tidal force follows $\ddot{H}_{\rm{m}}/\ddot{H}_{\rm{t}}\propto R^{1/2}$, meaning that as the gas approaches the black hole, the magnetic force cannot become dynamically important. This implies that the timescale $t^{\rm{hyd}}_{\rm{mag}}$ derived in equation~(\ref{eq:t_mag}) is valid in the bound part of the stream only as long as it is smaller than the time taken to reach the apocentre $t_{\rm{apo}}=\pi |2\mu|^{-3/2} t_{\rm{str}}(M_{\rm{h}}/M_{\star})^{1/2}$, which gives the following upper limit on the initial plasma beta
\begin{align}\label{eq:beta_crit_hd_bound}
    \beta_{\rm{M},\star} &\leq \beta^{\rm mag}_{\rm M,\star}\equiv\frac{\pi^{2}}{|2\mu|^{3}}\frac{M_{\rm{h}}}{M_{\star}}=\notag\\&=1.2 \times 10^{9} \left(\frac{0.1}{|\mu|}\right)^{3} \left(\frac{M_{\rm{h}}}{10^{6}\,M_{\odot}}\right) \left(\frac{M_{\star}}{M_{\odot}}\right)^{-1}.
\end{align}
Only if this condition is met can a bound stream element become magnetically dominated before its return to pericentre.

Similarly, in the unbound part of the stream that is in the hydrostatic regime ($0<\mu<\mu_{\rm{crit}}$), equations~(\ref{dotHp}) and~(\ref{dotHm}) hold as long as the stream elements are closer than the transition distance\footnote{This is where the gravitational potential due to the black hole becomes negligible, such that $v=\sqrt{2\,\epsilon}$. Inserting this into the condition $v^{2}/2 - GM_{\rm{h}}/R >0$ valid for unbound trajectories gives a lower constraint on the distance from the black hole: $R > R_{\rm{t}}\mu^{-1}(M_{\rm{h}}/M_{\star})^{1/3} \equiv R_{\rm{tr}}$ (note that this is the same as the apocentre distance, but for the opposite sign of $\mu$).} $R_{\rm{tr}}=R_{\rm{t}}\mu^{-1}(M_{\rm{h}}/M_{\star})^{1/3}$, that is where the stream elements reach their terminal velocity (see also \citet{Coughlin2016}). The stream reaches this distance after a time $t_{\rm{tr}}$ given approximately by
\begin{align}\label{eq:t_tr}
    t_{\rm{tr}} &= \frac{t_{\rm{str}}}{\mu^{3/2}} \left(\frac{M_{\rm{h}}}{M_{\star}}\right)^{1/2} \notag 
    \\
    &\approx 1.4 \times 10^{4}\, \rm{h} \, \left(\frac{0.1}{\mu}\right)^{3/2} \left(\frac{M_{\rm{h}}}{10^{6}\,M_{\odot}}\right)^{1/2} \left(\frac{R_{\star}}{R_{\odot}}\right)^{3/2} \left(\frac{M_{\star}}{M_{\odot}}\right)^{-1},
\end{align}
which is obtained using the scaling $t \propto R^{3/2}$ valid at early times for a nearly parabolic orbit. After the transition time $t_{\rm{tr}}$, the evolution of the elongation changes from $\ell \propto R^{2}$ to $\ell \propto R$ (see equation~(\ref{ell_propto_R})), and because the stream elements move at constant speed the distance to the black hole begins to follow $R \propto t$. If the hydrostatic equilibrium is maintained during this transition, then the width must follow $H \propto R^{1/4}$ (equation~(\ref{eq:H_hydro_late})).\footnote{This scaling was first found analytically by \citet{Kochanek1994} and later by \citet{Coughlin2016}, but we emphasize that it has not been proven numerically yet due to computational limitations.} 

Using these scalings, we find that here the gas pressure acceleration quickly drops as $\ddot{H}_{\rm{p}} \propto R^{-5/4} \propto t^{-5/4}$ and eventually gets surpassed by the magnetic one $\ddot{H}_{\rm{m}} \propto R^{1/4} \propto t^{1/4}$. This happens after a time $t_{\rm{mag}} = t_{\rm{tr}} \, \beta^{2/3}_{\rm{M,tr}}$, obtained using the scaling $\beta_{\rm{M}} \propto t^{-3/2}$. Using $\beta_{\rm{M}} \propto t^{-2}$ valid at earlier times, the plasma beta at time $t_{\rm{tr}}$ is $\beta_{\rm{M,tr}} = \beta_{\rm{M,\star}} (t_{\rm{str}}/t_{\rm{tr}})^{2}$. As a result, unbound stream elements for which $t^{\rm{hyd}}_{\rm{mag}} > t_{\rm tr}$ (equations~(\ref{eq:t_mag}) and~(\ref{eq:t_tr})) can still get magnetically supported later, after a time given by
\begin{equation}\label{eq:t_mag_tr}
    t^{\rm{hyd}}_{\rm{mag}} (\rm late)=t_{\rm{str}} \mu^{1/2} \beta^{2/3}_{\rm{M},\star} \left(\frac{M_{\rm{h}}}{M_{\star}}\right)^{-1/6}.
\end{equation}
The condition $t_{\rm mag}^{\rm hyd} (\rm late) > t_{\rm tr}$ corresponds to an upper limit on the initial plasma beta $\beta_{\rm{M},\star} \leq \mu^{-3}(M_{\rm{h}}/M_{\star})$, which is similar to that given by equation~(\ref{eq:beta_crit_hd_bound}) but for $\mu>0$. For a stream element with $\mu=0.1$, in the marginal case where $\beta_{\rm{M},\star} =10^{9}$, $t^{\rm{hyd}}_{\rm{mag}}(\rm early) \approx t^{\rm{hyd}}_{\rm{mag}}(\rm late)\approx 1.6\,$ years.

We note that these results were obtained assuming $\Gamma=5/3$ for the stellar profile and $\gamma=5/3$ for the adiabatic index. Both these parameters can play a role in the onset of magnetic pressure support. Varying $\Gamma$ changes the stellar density and pressure profiles, which in turn affects the initial plasma-beta $\beta_{\rm M,\star}$, and consequently the time $t^{\rm hyd}_{\rm mag}$ (see equation~(\ref{eq:t_mag})). For example, in a radiative polytrope ($\Gamma = 4/3$), the core is more centrally concentrated and pressurized, resulting in a higher $\beta_{M,\star}$ and $t^{\rm hyd}_{\rm mag}$ at fixed magnetic field strength compared to the $\Gamma = 5/3$ case considered in this work. Modifying the adiabatic index $\gamma$ can influence the evolution of the gas pressure acceleration $\ddot{ H}_{\rm p}$. Using the equation of state $P\propto \rho^{\gamma}$ we find that $\ddot{H}_{\rm p}  \propto t^{5/3-2\gamma}$ for an arbitrary $\gamma$. Comparing this with equation~(\ref{dotHm}), we conclude that the magnetic field can become dynamically important for any $\gamma>2/3$, a condition generically satisfied in astrophysical gases. As a result, the expansion of the stream due to magnetic pressure is expected to occur regardless of the choice of adiabatic index. However, this choice affects the time at which magnetic pressure becomes dominant,  with $t^{\rm hyd}_{\rm mag} \propto \beta^{3/(6\gamma-4)}_{\rm M, \star}$, which increases with decreasing $\gamma$.

\paragraph*{Ballistic regime ($|\mu|>\mu_{\rm crit}$).} In the ballistic regime the relevant forces to compare are the tidal force $\ddot{H}_{\rm{t}}$, which is the dominant one initially, and the magnetic force $\ddot{H}_{\rm{m}}$. In the right panel of Fig.~\ref{fig:Pg_Pm_time} we show their evolution for a bound and for an unbound stream element of boundness $|\mu|=1.1$, in the simulation initialized with a uniform magnetic field of strength $B_{\star}=10^{5} \, \rm{G}$. Since in this regime $H \propto R$ and $\ell \propto R^{2}$ initially (see equations~(\ref{eq:l_R2}) and~(\ref{eq:H_hydro_early})), at early times the magnetic acceleration follows $\ddot{H}_{\rm{m}} \propto R^{-1} \propto t^{-2/3}$ (see pink segment) and thus decreases slower than the tidal one $\ddot{H}_{\rm{t}} \propto R^{-2} \propto t^{-4/3}$ (see green segment). This means that the ratio
\begin{equation}
    \tilde{\beta}_{\rm{M}}\equiv\ddot{H}_{\rm{t}}/\ddot{H}_{\rm{m}}
\end{equation}
can eventually reach unity, making the magnetic field dynamically important also in this region of the stream. Since $\tilde{\beta}_{\rm{M}}\propto R^{-1}\propto t^{-2/3}$, this will happen at a time 
\begin{equation}\label{eq:t_mag_bal}
    t^{\rm{bal}}_{\rm{mag}} = t_{\rm{str}}\tilde{\beta}^{3/2}_{\rm{M},\star}
\end{equation}
after disruption. 
\begin{figure*}
    \includegraphics[width=1\textwidth]{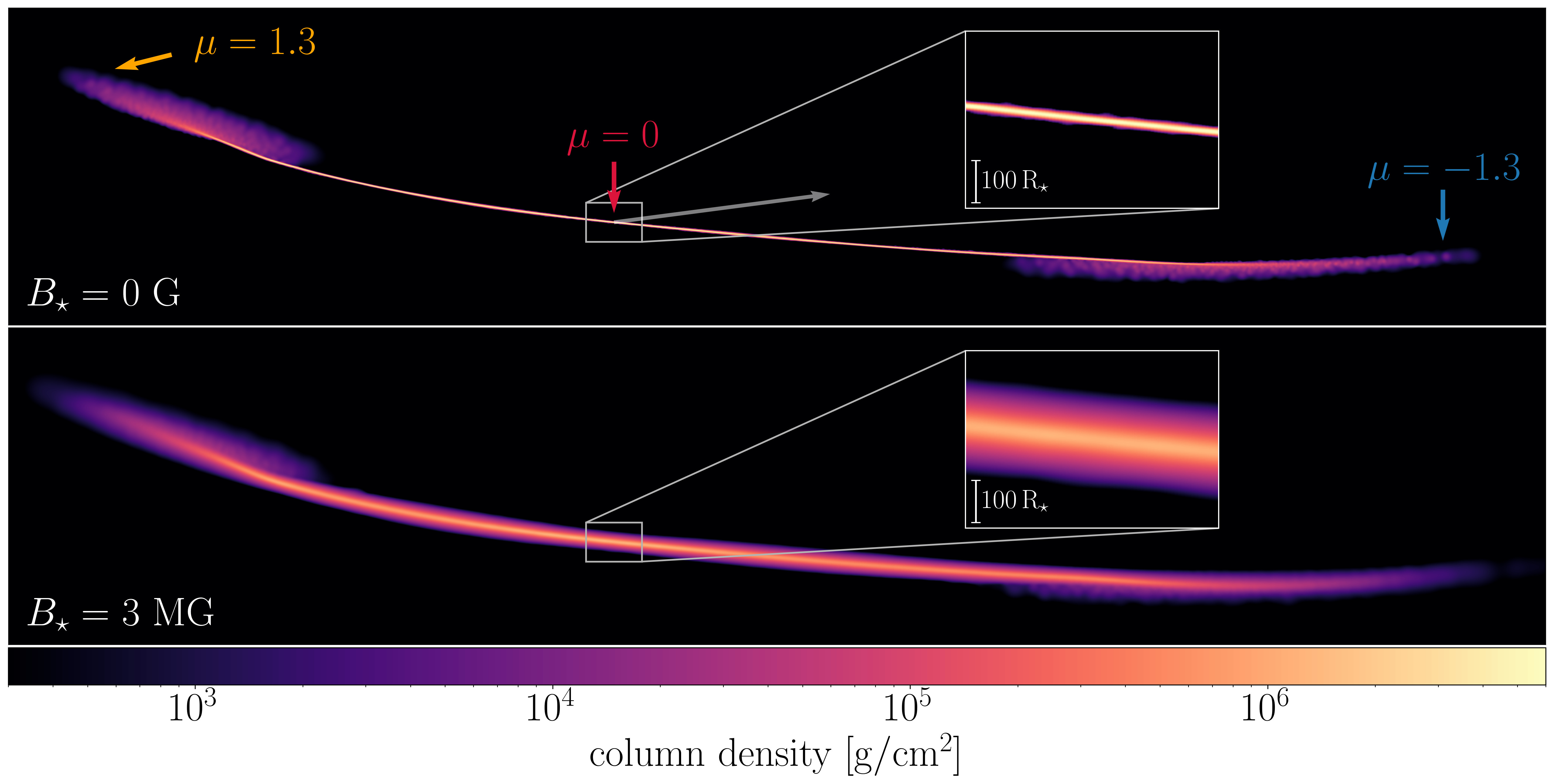}
    \caption{Projection of the density on the orbital plane for the hydrodynamic (top) and magnetized (bottom) tidal stream with initial field $B_{\star}=3\,\rm MG$ at time $t\approx 442\, \rm{h}$. The grey arrow points towards the black hole. The inset plot shows a zoom-in of the near-parabolic stream section ($|\mu|<0.05$), which at this point of the simulation is located at $163 \, R_{\rm{t}}$ from the black hole. The width of the magnetized stream here is a factor of a few larger than in the hydrodynamic case.}
    \label{fig:density_maps}
\end{figure*}
However, the magnetic field can only become dynamically important before $t_{\rm{tr}}$ for unbound stream elements in the ballistic regime. This is because for $t>t_{\rm tr}$, the elongation follows $\ell \propto R \propto t$, such that $\tilde{\beta}_{\rm{M}}=\rm{const}$. This effect is visible in our simulation with $B_{\star}=10^5 \, \rm{G}$ (see right panel of Fig.~{\ref{fig:Pg_Pm_time}}). Here, the initial ratio between the tidal and the magnetic acceleration is $\tilde{\beta}_{\rm{M},\star}\approx 10^{4}$, which gives $t^{\rm{bal}}_{\rm{mag}}\approx 10^5 \, \rm{hours}$. This is much greater than the transition time $t_{\rm{tr}}\approx 10^2 \, \rm{hours}$. As expected, we find that in this case the magnetic acceleration remains lower than the tidal one and for $t>t_{\rm tr}$ the two forces begin to evolve similarly as $\ddot{H}_{\rm{m}}\propto \ddot{H}_{\rm{t}} \propto R^{-2} \propto t^{-2}$ (see yellow segments in the right panel of Fig.~{\ref{fig:Pg_Pm_time}}), such that the magnetic field cannot dominate the dynamics hereafter. Using equation~(\ref{eq:t_mag_bal}) and setting $t^{\rm{bal}}_{\rm{mag}}<t_{\rm{tr}}$, we find that the unbound part of the stream that is initially in the ballistic regime can become magnetically supported only if 
\begin{equation}
    \tilde{\beta}_{\rm{M},\star}<\frac{1}{\mu} \left(\frac{M_{\rm{h}}}{M_{\star}}\right)^{1/3}=\frac{100}{\mu}\left(\frac{M_{\rm{h}}}{10^6\,M_{\odot}}\right)^{1/3}.
\end{equation}
For a stream element with $\mu=1$, in the limiting case where $\tilde{\beta}_{\rm{M},\star}= 100$, the magnetization timescale $t^{\rm{bal}}_{\rm{mag}}$ equals the transition time $t_{\rm{tr}}$ that is roughly $\approx 442 \, \rm{hours}$ ($\approx 18 \, \rm{days}$).

Similarly to the unbound debris, in the bound part of the stream that belongs to the ballistic regime, equation~(\ref{eq:t_mag_bal}) applies before the apocentre, which corresponds to $\tilde{\beta}_{\rm{M},\star}\leq \pi^{2/3}|2\mu|^{-1} (M_{\rm{h}}/M_{\star})^{1/3}$, as obtained by setting $t^{\rm{bal}}_{\rm{mag}}<t_{\rm{apo}}$. For higher values of $\tilde{\beta}_{\rm{M},\star}$, however, the magnetic field can still become dynamically important as the stream infalls towards the black hole. This effect is visible in our simulation with $B_{\star}=10^5\,\rm{G}$, where for a stream section with $\mu=-1.1$, the magnetic acceleration exceeds the tidal force at a time $t>t_{\rm{apo}}$ (see grey lines in the right panel of Fig.~\ref{fig:Pg_Pm_time}). This occurs because as the evolution of the elongation changes to $\ell \propto R^{-1/2}$ (see equation~(\ref{eq:ell_propto_R-12})), $\tilde{\beta}_{\rm{M}}\propto R^{3/2}$, meaning that as the distance to the black hole decreases, the magnetic force can eventually dominate over the tidal force. This happens at a distance from the black hole given by $R^{\rm bal}_{\rm mag}=R_{\rm{apo}}\tilde{\beta}^{-2/3}_{\rm{M,apo}}$, where $\tilde{\beta}_{\rm{M,apo}}$ is the ratio between tidal and magnetic force at $R_{\rm{apo}}$. Using the scaling $\tilde{\beta}_{\rm{M}}\propto R^{-1}$ valid before $t_{\rm{apo}}$, $\tilde{\beta}_{\rm{M,apo}}=\tilde{\beta}_{\rm{M},\star}(R_{\rm{t}}/R_{\rm{apo}})$, and therefore $R^{\rm bal}_{\rm mag}=R_{\rm{t}}|\mu|^{-5/3}\tilde{\beta}^{-2/3}_{\rm{M},\star}(M_{\rm{h}}/M_{\star})^{5/9}$. In order for the magnetic field to become dynamically important before the second pericentre passage $R^{\rm bal}_{\rm mag}$ must be greater than the tidal radius $R_{\rm{t}}$, which gives the following condition for the initial ratio between tidal and magnetic forces:
\begin{equation}
    \tilde{\beta}_{\rm{M},\star}<\frac{1}{|\mu|^{5/2}}\left(\frac{M_{\rm{h}}}{M_{\star}}\right)^{5/6}.
\end{equation}
This implies, for example, that a very bound stream element with $\mu=-1$ will become magnetically supported while it infalls towards the black hole provided that $10^2\lesssim\tilde{\beta}_{\rm{M},\star}\lesssim10^{5}$.

\paragraph*{Fraction of mass that can become magnetized.} Combining equations~(\ref{eq:t_mag}),~(\ref{eq:t_mag_tr}) and~(\ref{eq:t_mag_bal}), together with the corresponding conditions on $\beta_{\rm{M},\star}$ and $\tilde{\beta}_{\rm{M},\star}$ discussed above, we obtain an expression for the time $t_{\rm{mag}}(\mu,B_{\star})$ after which the magnetic field becomes important for a given $\mu$, as a function of the stellar magnetic field $B_{\star}$. We can now use this time to estimate the mass 
\begin{equation}\label{M_mag}
    M_{\rm{mag}}(t_{\rm{target}},B_{\star}) = R_{\star} \int_{t_{\rm{mag}}(\mu,B_{\star}) < t_{\rm{target}}}\Lambda(\mu)d\mu,
\end{equation}
of the stream that becomes magnetically supported within a time $t_{\rm{target}}$ after disruption for a given initial magnetic field $B_{\star}$. Computing $t_{\rm{mag}}(\mu,B_{\star})$ requires to know how $\beta_{\rm{M},\star}$ and $\tilde{\beta}_{\rm{M},\star}$ relate to $\mu$ for a given stellar magnetic field $B_{\star}$. To accurately link these quantities we make use of our numerical simulation. We consider a snapshot of the simulation initialized with a constant magnetic field $B_{\star}=1\,\rm{G}$ at a time $t =3.5\,\rm days$, when the orbital energy distribution in the stream has settled. We slice the stream into thin sections based on $\mu$ and for each of them we use the properties that they had at $t=0$ to evaluate $\beta_{\rm{M},\star}$ and $\tilde{\beta}_{\rm{M},\star}$ through a mass average. Since both quantities are inversely proportional to magnetic pressure, we can simply rescale them by $(1\,\rm{G}/B_{\star})^{2}$ to obtain $\beta_{\rm{M},\star}$ and $\tilde{\beta}_{\rm{M},\star}$ for any $B_{\star}$. We then use this to obtain $t_{\rm{mag}}(\mu,B_{\star})$ along the stream\footnote{To determine whether a stream section is in the hydrostatic or ballistic regime in the simulation, we verify whether its density is, respectively, above or below the critical value $\rho_{\rm crit}=M_{\rm h}/(2\pi R^3)$ separating the two regimes.} and thus, calculate the magnetized mass given by equation~(\ref{M_mag}) as a function of stellar magnetic field $B_{\star}$. 

The result is shown in Fig.~\ref{fig:mass_fraction}. For $t_{\rm{target}}=+\infty$ (grey solid line), roughly half of the stream, corresponding to the unbound debris in the hydrostatic regime, becomes magnetically supported for any $B_{\star}>0$, as shown by the plateau towards low $B_{\star}$. This is because, as found in the calculations above, in this part of the stream magnetic pressure has an infinite amount of time to become dynamically important. Instead, for the bound debris there is the additional constraint that the magnetic field has to become important before reaching apocentre, which is why higher field strengths are needed in this case (see equation~(\ref{eq:beta_crit_hd_bound})). For $B_{\star} \gtrsim 10^5\, \rm G$, the entire bound debris becomes magnetically supported. Finally, for $B_{\star}\gtrsim 10^6 \, \rm G$, the fraction of magnetized mass undergoes a last increase associated with the unbound debris in the ballistic regime. When we require the magnetic field to become dynamically important in a finite time after disruption (dashed and solid black lines), the minimum magnetic field needed to cause a dynamical effect is shifted towards higher values. In particular, we find that, in order for the magnetic field to become dynamically important in at least $\approx 50\,\%$ of the mass of the stream within $1$ year after disruption, the initial magnetic field in the star must be at least $\sim 10^4\,\rm{G}$.

\subsubsection{Stream width increase under magnetic pressure}\label{Ss:Stream width increase under magnetic pressure}
\begin{figure}
    \includegraphics[width=0.98\columnwidth]{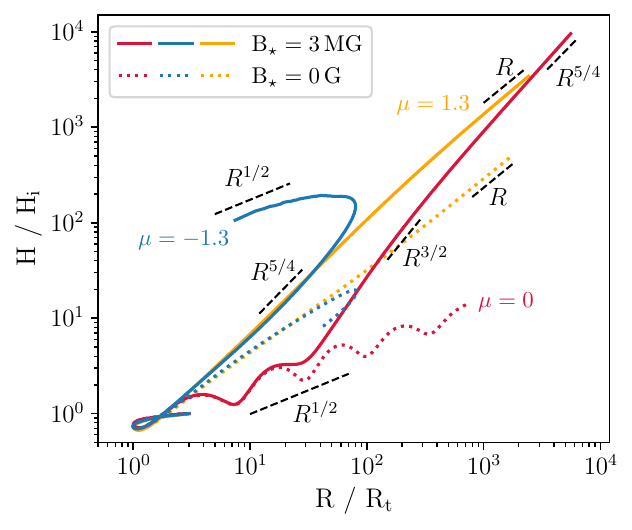}
    \caption{Evolution of the stream vertical width as a function of the distance from the black hole in the simulation for different orbital energies corresponding to $\mu=-1.3$ (blue lines), $\mu=0$ (red lines), and $\mu=1.3$ (yellow lines). The dashed lines correspond to a simulation without magnetic fields, while the solid lines correspond to magnetized cases with initial magnetic field $B_{\star}=3\,\rm MG$. The black dashed segments indicate the expected scalings for the different parts of the evolution.}
    \label{fig:H_vs_R}
\end{figure}
As described in the previous section, the stellar magnetic field can eventually become dynamically important during a TDE, providing additional support along the transverse size of the stream through the action of magnetic pressure. This effect can be seen in Fig.~\ref{fig:density_maps}, which shows a column density map of the stream at time $t = 442 \, \rm{h}$, in the hydrodynamic simulation (top panel) and in the MHD simulation initialized with a uniform field $B_{\star}=3\,\rm{MG}$ (bottom panel). Here $t>t_{\rm mag}\approx 44\,\rm hours$ for the entire stream (see Fig.~\ref{fig:Pg_Pm_time}), such that magnetic pressure becomes dynamically important, causing the stream to become wider in the transverse direction by a factor of a few.

To quantify this effect, we measure the local thickness in these simulations as described in Section~\ref{ss:Onset of magnetic pressure support}. We show the resulting vertical width as a function of the distance from the black hole in Fig.~\ref{fig:H_vs_R} for different stream elements and we compare the non-magnetized and magnetized cases. In the absence of magnetic fields, the width of the parabolic section with $\mu=0$ oscillates around the scaling $H \propto R^{1/2}$ (see dotted red line), as expected from hydrostatic equilibrium. When including magnetic fields (solid red line), the evolution is drastically modified once magnetic pressure becomes significant in the stream. Here the vertical acceleration reduces to $\ddot{H} \approx \ddot{H}_{\rm{m}} \approx B^{2}/(8\pi\rho H)$, such that the stream begins to expand at speed $\dot{H}\approx v_{A}(t_{\rm{mag}})$\footnote{Since $\ddot{H}\approx B^{2}/(8\pi\rho H)$, the velocity at which the stream expands can be approximated as $\dot{H}\sim \ddot{H}\, t \approx B/\sqrt{8\pi\rho}$.}, where $v_{A}(t_{\rm{mag}})=B(t_{\rm{mag}})/\sqrt{4\pi\rho(t_{\rm{mag}})}$ is the Alfv\'en velocity at time $t_{\rm{mag}}$. As shown in Fig.~\ref{fig:Pg_Pm_time}, because of this expansion the magnetic acceleration starts to decline. Consequently, the stream begins to expand at constant speed, with its width scaling as $H\propto t \propto R^{3/2}$ (see black dashed segment in Fig.~\ref{fig:H_vs_R}). During this free expansion the tidal acceleration $\ddot{H}_{\rm{t}}\propto t^{-1}$ decreases slower than the magnetic one $\ddot{H}_{\rm{m}}\propto t^{-5/3}$ (see yellow dashed segments in Fig.~\ref{fig:Pg_Pm_time}), such that $\ddot{H}_{\rm{m}}$ and $\ddot{H}_{\rm{t}}$ become comparable in a finite time $t_{\rm{eq}}$. For $t>t_{\rm{mag}}$, $\tilde{\beta}_{\rm{M}}=\ddot{H}_{\rm{t}}/\ddot{H}_{\rm{m}} \propto t^{2/3}$, implying that $t_{\rm{eq}}=t_{\rm{mag}}\tilde{\beta}^{-3/2}_{\rm{M},\rm{mag}}$, where $\tilde{\beta}_{\rm{M,mag}}$ is the ratio between tidal and magnetic forces at time $t_{\rm{mag}}$. Since $\tilde{\beta}_{\rm{M}}\propto t^{-2}$ at $t<t_{\rm{mag}}$, $\tilde{\beta}_{\rm{M,mag}}=\tilde{\beta}_{\rm{M},\star}/\beta_{\rm{M},\star}$, which leads to $t_{\rm{eq}}=t_{\rm{mag}}(\beta_{\rm{M},\star}/\tilde{\beta}_{\rm{M},\star})^{3/2}$. Using $t_{\rm{mag}}\approx 44 \, \rm{hours}$ and $\beta_{\rm{M},\star}/\tilde{\beta}_{\rm{M},\star}=\ddot{H}_{\rm{p},\star}/\ddot{H}_{\rm{t},\star} \approx 10$ (as found in our simulation, see Fig.~\ref{fig:Pg_Pm_time}), $t_{\rm{eq}}\approx 1320 \, \rm{hours}$, which is consistent with the time after which $\ddot{H}_{\rm{t}}=\ddot{H}_{\rm{m}}$ in our simulation (see rightmost red vertical line in the left panel of Fig.~\ref{fig:Pg_Pm_time}). 

After $t_{\rm{eq}}$ the free expansion phase stops and the subsequent evolution of the parabolic stream element is driven by a quasi-equilibrium between the outward magnetic force and the inward tidal force, leading to $\ddot{H}_{\rm{m}}\propto \ddot{H}_{\rm{t}}$. By plugging $\ddot{H}_{\rm{t}}=GM_{\rm h}H/R^3$ and $\ddot{H}_{\rm{m}}=B^2/(\rho H)$ into this proportionality relation, and using magnetic flux conservation, we find that during this equilibrium phase the width is expected to follow
\begin{equation}\label{eq:H_R54}
    H \propto \ell^{1/4} R^{3/4}.
\end{equation}
By inserting the scaling $\ell \propto R^{2}$ valid for the parabolic stream element, we find $H \propto R^{5/4}$, which is in good agreement with the late-time width increase shown in Fig.~\ref{fig:H_vs_R} for $\mu=0$ (see black dashed segment below the solid red line).\footnote{Since $H \propto R^{5/4} \propto t^{5/6}$ during the equilibrium between tidal and magnetic force, the velocity at which the stream expands follows $\dot{H} \propto t^{-1/6}$. As a result $\ddot{H}=\ddot{H}_{\rm{m}}+\ddot{H}_{\rm{t}}<0$, meaning that during this equilibrium the tidal force slightly exceeds the magnetic one.} 

The stability of this equilibrium between the magnetic force and the tidal force can be understood by imposing the stream to be slightly over-compressed at increasing radii, with $H\propto R^{5/4-\eta}$ and $0<\eta \ll 1$. This results in $\ddot{H}_{\rm{m}} / \ddot{H}_{\rm{t}} \propto R^{4\eta}$, implying that the repulsive magnetic force increases to restore the equilibrium. This equilibrium is therefore stable and we expect the width of the parabolic stream element to keep expanding indefinitely following equation~(\ref{eq:H_R54}). This shows that, in contrast with the hydrodynamic evolution, the bulk of a magnetized stream element is not confined by self-gravity for $t>t_{\rm{mag}}$.  

We now describe the effect of magnetic pressure on stream elements that are initially in the ballistic regime. In Fig.~\ref{fig:H_vs_R} we show the width evolution for a very bound and a very unbound stream section with $|\mu|\pm1.3$. In the absence of magnetic fields, the thickness follows $H \propto R$ (see dotted blue and yellow lines), as expected from ballistic motion. When including magnetic fields, once these become dynamically important the evolution becomes similar to the parabolic case, showing a fast width increase. However, because the tidal force dominates the dynamics from the start, the free expansion is absent here, and the width follows $H\propto R^{5/4}$ as soon as the magnetic field becomes significant (see solid yellow and blue lines in Fig.~\ref{fig:H_vs_R}). Later, at times $t>t_{\rm{tr}}\approx 380 \, \rm h$, in the unbound part of the stream the elongation starts to grow slower, $\ell \propto R$, implying that the width scaling transitions to $H \propto R $, according to equation~(\ref{eq:H_R54}), corresponding to another stable equilibrium. This is in good agreement with the width increase shown in Fig.~\ref{fig:H_vs_R} (see solid yellow line). This transition eventually occurs also in the unbound part of the stream that is initially in the hydrostatic regime. We note that although this late-time scaling for the unbound gas in the ballistic regime is linear like in the non-magnetized case, at late times the width can nevertheless be significantly larger due to the steeper scaling at early times when $t<t_{\rm{tr}}$. 

While moving away from the black hole, bound stream elements evolve analogously to the unbound ones, showing the width increase $H \propto R^{5/4}$ expected when magnetic force and tidal force are in equilibrium. This is shown by the solid blue line in Fig.~\ref{fig:H_vs_R}, that describes the width evolution for a bound stream element with $\mu=-1.3$ in our simulation. This equilibrium is perturbed at apocentre, where the evolution of the elongation changes from $\ell \propto R^{2}$ to $\ell \propto R^{-1/2}$ (see equation~(\ref{eq:ell_propto_R-12})). Inserting this new scaling into equation~(\ref{eq:H_R54}) shows that, in order for this equilibrium to be maintained, the width should follow $H\propto R^{5/8}$ during the infall. With this width evolution, however, such equilibrium would cause a stronger compression of the infalling stream than that caused by the tidal force only ($H \propto R^{1/2}$). But since the tidal and the magnetic force act in opposite directions along the transverse size of the stream, this behaviour is unphysical, and the equilibrium must therefore cease at apocentre. During the infall, the tidal force instead dominates the dynamics, causing the stream width to decrease following $H \propto R^{1/2}$, consistently with the width evolution found in our simulation (see blue solid line in Fig.~\ref{fig:H_vs_R}). 

As described in Section~\ref{ss:Onset of magnetic pressure support}, if the magnetic field does not become significant while moving away from the black hole, it can still start dominating the dynamics during the infall if $10^2\lesssim\tilde{\beta}_{\rm{M},\star}\lesssim10^{5}$ in parts of the stream that are initially dominated by the tidal force. If this happens, the transverse evolution of the infalling stream element is driven by the equilibrium between the tidal and the magnetic force, which in this case is physically allowed, since the compression caused by this equilibrium leads to a scaling $H \propto R^{5/8}$, which is weaker than that caused by the tidal force only ($H \propto R$). Although the compression in this case is driven by a different mechanism, the resulting thickness evolution is very similar to the case where the magnetic field becomes important before apocentre. Furthermore, this effect occurs only in a small fraction of the stream mass and within a narrow range of initial field strengths.

\subsubsection{Width of the returning stream}
\begin{figure}
\includegraphics[width=0.98\columnwidth]{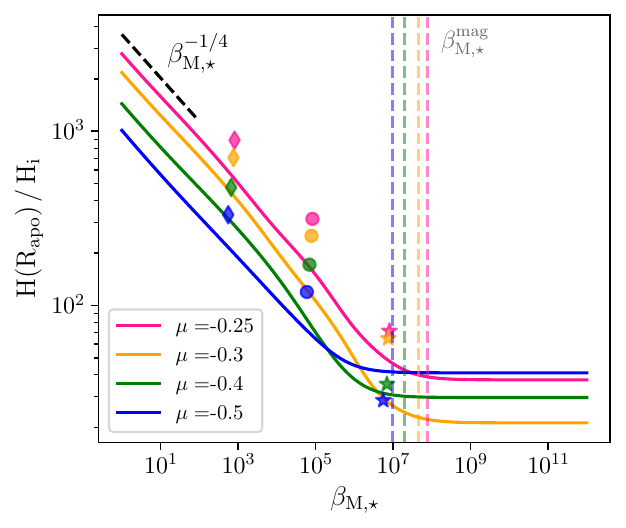}
    \caption{Stream width evaluated at apocentre vs initial plasma beta for different bound stream elements initially in the hydrostatic regime. The markers correspond to the result of three simulations with a uniform magnetic field of strength $B_{\star}=10^{5} \, \rm{G}$ (stars), $B_{\star}=10^{6} \, \rm{G}$ (points) and $B_{\star}=10^{7} \, \rm{G}$ (diamonds), while the bold lines are obtained with the semi-analytic model. Different colours correspond to different stream sections. The black dashed segment indicates the expected analytic scaling $\beta^{-1/4}_{\star}$ (see equation~(\ref{eq:beta-14})). The coloured dashed lines indicate the critical $\beta_{\rm{M},\star}$ below which the magnetic field is expected to become dynamically important and they are obtained using equation~(\ref{eq:beta_crit_hd_bound}).}
    \label{fig:H_vs_beta}
\end{figure}
The width of the returning stream is of crucial importance in the later phases of a TDE, e.g. the nozzle shock, the stream self-intersection and the formation of the accretion disk. We investigate its dependence on the initial magnetic field strength by measuring the width in three numerical simulations initialized with a uniform magnetic field of strength $B_{\star} = 10^{5},10^{6}$ and $10^{7} \,\rm{G}$. We also employ the semi-analytic model presented in Section~\ref{sec:Semi-analytic model of a magnetized TDE}, which enables us to explore a wider region of the parameter space. In the results presented here and throughout the paper, the semi-analytic model assumes the magnetic field to be initially aligned with the stream’s elongation. This is justified by our finding that the results are insensitive to the initial field orientation, as expected from the rapid alignment discussed in Section~\ref{sec:Magnetic field re-alignment}.

With both methods, we evaluate for different stream elements the width at apocentre $H(R_{\rm{apo}})$\footnote{We evaluate the stream width at apocenter instead of closer to the black hole because of the lower numerical resolution during infall, which causes this part of the evolution to be less accurate.} as a function of the initial plasma beta $\beta_{\rm{M},\star}$. The result is shown in Fig.~\ref{fig:H_vs_beta} for different orbital energies $\mu$. The lines indicate the result of the semi-analytic model, while the markers correspond to the result of the simulation. For the latter, the initial plasma beta is obtained by averaging over the particles belonging to the section. In every slice considered in Fig.~\ref{fig:H_vs_beta} the $1\sigma$ uncertainty on the initial plasma beta is $\approx 0.7<\beta_{\rm{M},\star}>$, where $<.>$ is a mass average. We also note that, because the orbital energy distribution in the numerical simulation is wider than in the semi-analytic model, also the absolute value of the critical orbital energy separating the hydrostatic and ballistic regimes is higher. To circumvent this discrepancy, we restrict this analysis to stream sections with boundness lower than $|\mu|=0.533$, which corresponds to the critical boundness in the semi-analytic model. This ensures that the stream sections considered are initially in the same regime (the hydrostatic one) both in the simulation and in the semi-analytic model. 

As shown by the solid lines in Fig.~\ref{fig:H_vs_beta}, in the limit of high $\beta_{\rm{M},\star}$ the magnetic field does not become dynamically important and the final stream width $H(R_{\rm{apo}})$ is therefore independent of the initial plasma beta $\beta_{\rm{M},\star}$. There is a critical $\beta_{\rm{M},\star}$ below which magnetic pressure causes the width to deviate from the hydrodynamic evolution. The threshold obtained with the semi-analytic model (see solid lines) is consistent with our estimate of the maximum plasma beta $\beta^{\rm{mag}}_{\rm{M},\star}$ (dashed vertical lines) required initially for the magnetic field to become dynamically important, which is given by equation~(\ref{eq:beta_crit_hd_bound}). Below this critical value, the width at apocentre and the initial plasma beta are related by a simple power law (see dashed black segment), which can be understood as follows. Using the scaling $H\propto R^{5/4}$ (see e.g. red solid line in Fig.~(\ref{fig:H_vs_R})), for a stream section affected by magnetic fields $H(R_{\rm{apo}}) \propto H(R^{\rm hyd}_{\rm{mag}}) \left(R^{\rm hyd}_{\rm{mag}}\right)^{-5/4}$. Since $H \propto R^{1/2}$ at $R < R^{\rm hyd}_{\rm{mag}}$ and $R^{\rm hyd}_{\rm{mag}} \propto \beta^{1/3}_{\rm M,\star}$,
\begin{equation}\label{eq:beta-14}
    H(R_{\rm{apo}}) \propto \left(R^{\rm hyd}_{\rm{mag}}\right)^{-3/4} \propto \beta^{-1/4}_{\rm M,\star}.
\end{equation}
Because during infall the evolution of the width is not affected by magnetic fields, we expect equation~(\ref{eq:beta-14}) to be preserved until the first passage of the stream to pericentre. As shown in Fig.~\ref{fig:H_vs_beta}, because of this scaling the stream thickness can increase by up to a factor of $\sim 100$, while keeping the original star non-magnetized ($\beta_{\rm M,\star}>1$).

As shown in Fig.~\ref{fig:H_vs_beta}, both the semi-analytic model and the simulation are able to recover the expected scaling $H \propto \beta^{-1/4}_{\rm M,\star}$.
While this qualitative behaviour is well captured, the simulation and the semi-analytic model differ by a factor of a few for the exact normalization of the stream width. One potential contributing factor is given by the hydrodynamic oscillations that are exited during the first pericentre passage (see red dotted line in Fig.~\ref{fig:H_vs_R}). These have a role in determining the critical plasma beta below which the magnetic field can become significant, since they can anticipate or delay the emergence of magnetic pressure support. Because the phase and amplitude of these oscillations can differ between model and simulation due to the simplified treatment used in the former, we do expect some differences between the two methods.

\begin{figure}
\includegraphics[width=1\linewidth]{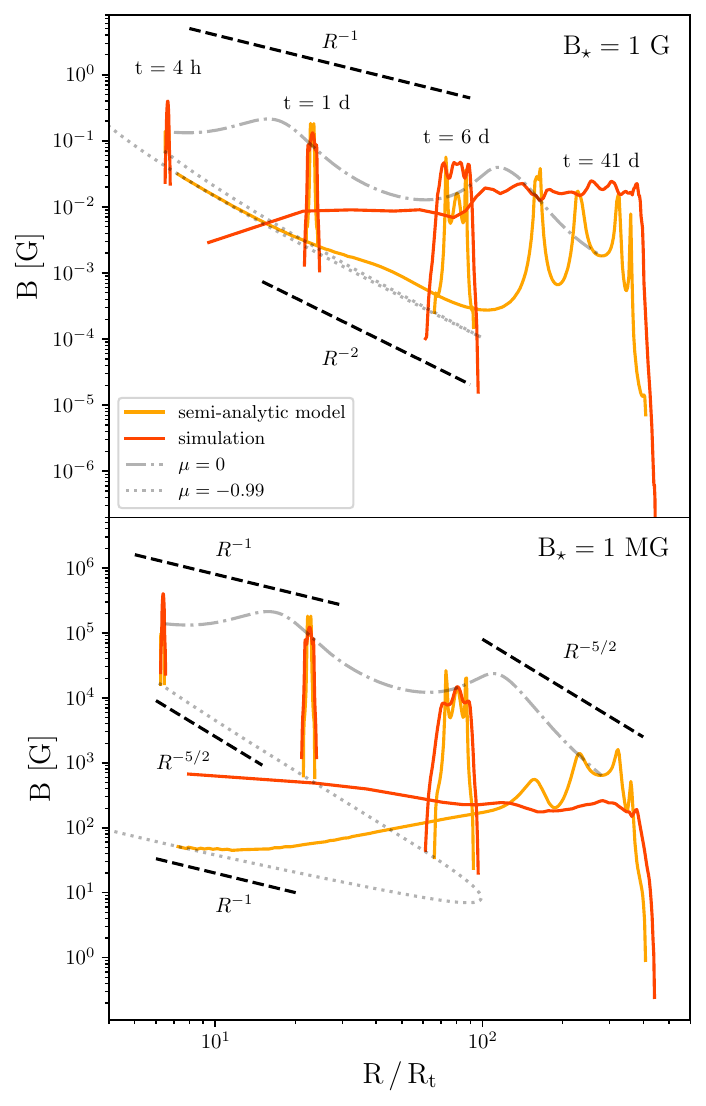}
    \caption{Magnetic field strength along the stream as a function of the distance from the black hole at different times, in the case of a magnetically dominated (bottom panel) and non dominated (top panel) stream. The coloured lines are obtained with the semi-analytic model (yellow) and the simulation (red). The grey lines are the paths followed by the magnetic field of two specific stream elements with boundness $\mu=0$ and $\mu=-0.99$. The black dashed segments indicate the expected scalings.}
    \label{fig:HBdistr}
\end{figure}
\subsection{MHD properties of TDE streams}\label{S:MHD properties of TDE streams}

\subsubsection{Magnetic field evolution along the stream}\label{Magnetic field evolution along the stream}
Regardless of its dynamical impact on the stream, the stellar magnetic field can undergo substantial changes during the tidal disruption and the subsequent evolution of the stream around the black hole. In this section we investigate the temporal evolution of the strength and topology of the magnetic field along the stream, before it returns to pericentre.

In Fig.~\ref{fig:HBdistr} we show the magnetic field strength along the stream as a function of the distance from the black hole at different times ($t = 4 \, \rm{h}, 1 \, \rm{d}, 6 \, \rm{d}$ and $41 \, \rm{d}$ after the stellar disruption), where the latest time corresponds to the rightmost curve. In the top panel we show a weakly magnetized case ($B_{\star}=1 \, \rm{G}$) where the magnetic field does not become dynamically important, while the bottom panel shows a magnetically dominated case ($B_{\star}=1 \, \rm{MG}$). The yellow lines are obtained using the semi-analytic model, while the red ones are the result of the simulation. The grey dotted and dash-dotted lines show, respectively, the evolution of the magnetic field for a fluid element in the ballistic regime ($\mu=-0.99$), and one in the hydrostatic regime ($\mu=0$), and they are obtained with the semi-analytic model. 

As a result of magnetic flux conservation, the evolution and the shape of the magnetic field distributions shown in Fig.~\ref{fig:HBdistr} are entirely determined by the evolution of the width. In the case of a very weak stellar magnetic field (top panel), which does not become dynamically important, the width follows $H \propto R^{1/2}$ in the hydrostatic regime and $H \propto R$ in the ballistic regime. This, together with flux conservation, causes the magnetic field to drop as $B \propto H^{-2} \propto R^{-1}$ and $\propto R^{-2}$, respectively, in the two regimes. These different scalings (see black dashed segments) explain why, as the stream moves away from the black hole, at the edges the magnetic field eventually becomes much weaker than in the bulk of the stream. Furthermore, the hydrodynamic oscillations affecting the stream elements that are initially in the hydrostatic regime, are imprinted on the magnetic field profiles (see dash-dotted line in the top panel). Specifically, because the phase and amplitude of these width oscillations vary along the stream, at a given time, some stream elements may be compressing while others are expanding. This gives rise to the oscillatory shape of the magnetic field distribution in the bulk of the stream. 
\begin{figure}
\includegraphics[width=1\linewidth]{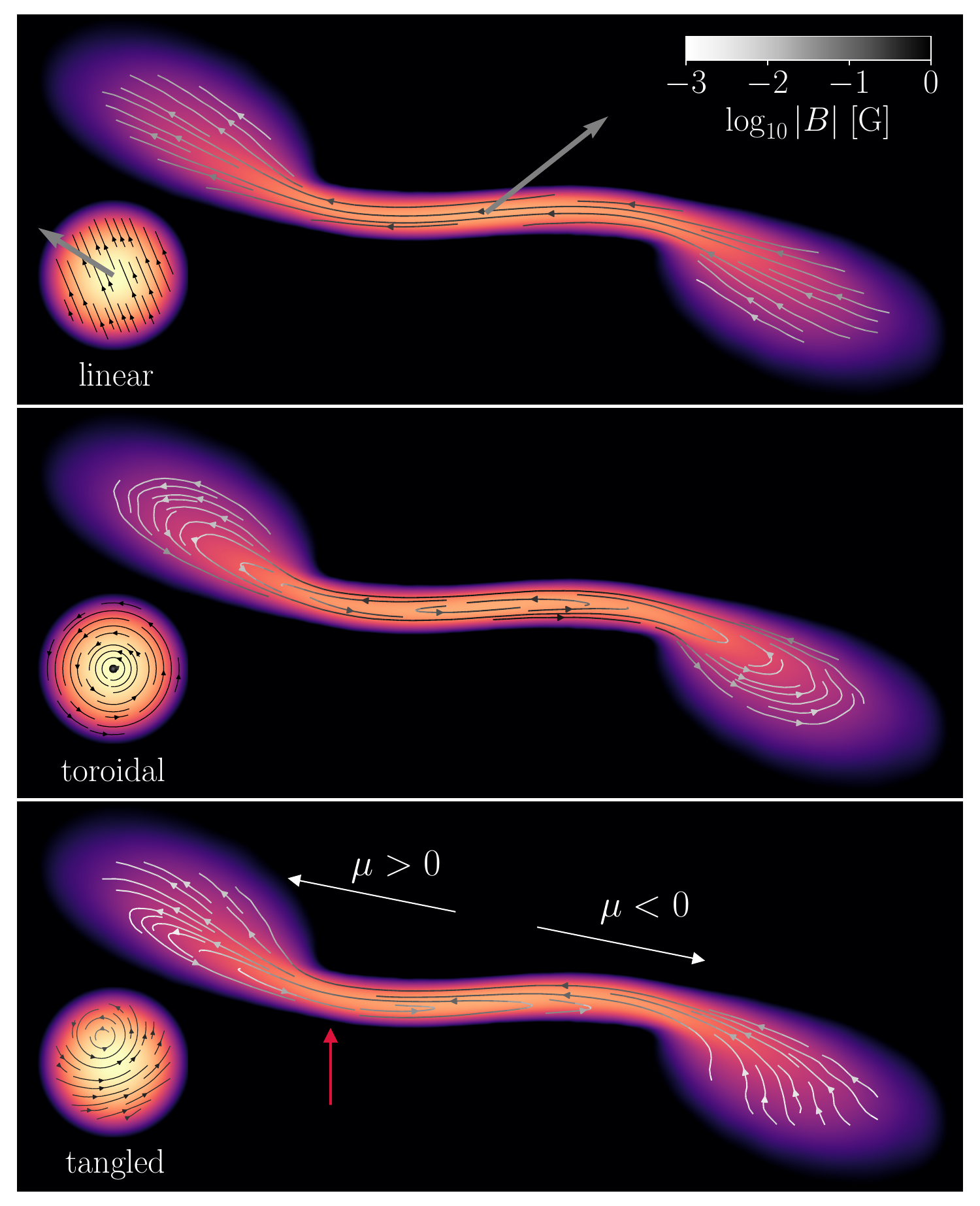}
    \caption{Density maps showing the magnetic field lines in the orbital plane at time $t = 7.96 \, \rm{h}$ for three numerical simulations initialized with a uniform (top), toroidal (centre) and tangled (bottom) magnetic field. The insets show the initial magnetic field in the star when located at $R=3 \, R_{\rm t}$. The colour bar shows the $\log_{10}|B|$. The gray arrows point towards the black hole, while the red arrow in the bottom panel indicates the region of high magnetic curvature in the tangled case.}
    \label{fig:B_lines_geometries}
\end{figure}
\begin{figure}\hspace{-0.4cm}
\includegraphics[width=1.05\linewidth]
{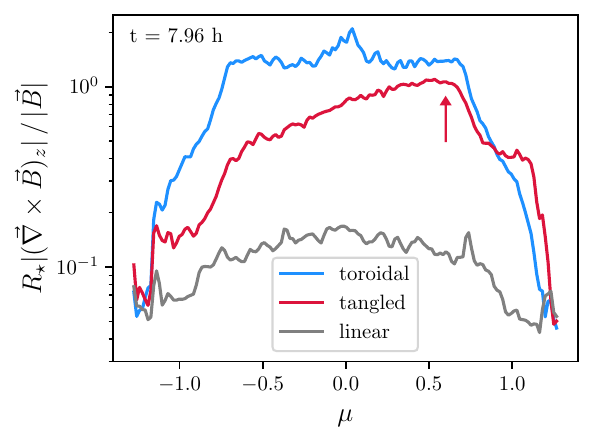}
    \caption{Normalized curl of the magnetic field as a function of $\mu$ at time $t=7.96\,\rm{h}$, for three different magnetic field configurations: uniform (grey line), toroidal (blue line) and tangled (red line). The red arrow indicates the peak of the curl in the tangled case, which corresponds to the region of high field curvature in the stream (see red arrow in Fig.~\ref{fig:B_lines_geometries}).}
    \label{fig:Curl_of_B}
\end{figure}
Another important feature in the magnetic field profile is visible at time $t=41\,\rm{days}$, when the most bound debris has returned to pericentre. Since this part of the stream is initially dominated by the tidal force, during infall its width follows $H \propto R$, causing the magnetic field to grow following $B \propto R^{-2}$. Some discrepancies between the semi-analytic model and the simulation become apparent at this late time for the part of the stream at low radii. Here, the simulation becomes less accurate because of the poorer resolution. In addition, some differences in the profile are expected because the orbital energy distribution in the simulation is broader compared to the semi-analytic model. As a result, a stream element of a given orbital energy may be in different physical regimes in the simulation and in the model, leading to distinct scaling behaviours.

In the case of a dynamically important magnetic field (bottom panel of Fig.~\ref{fig:HBdistr}), the width eventually deviates from the hydrodynamic evolution and begins to quickly grow following $H \propto R^{5/4}$ (see Section~\ref{Ss:Stream width increase under magnetic pressure}). Consequently, the magnetic field intensity follows $B \propto R^{-5/2}$, which is in good agreement with our findings. In this strongly magnetized case, the hydrodynamic equilibrium ceases eventually, such that the oscillations described above are quickly damped, resulting in a smoother magnetic field profile. Furthermore, during infall, the magnetic field amplification due to the stream getting compressed is much weaker. This is because, in contrast to the stronger compression as $H \propto R$ expected for the most bound debris in the ballistic regime, in the magnetically dominated case the width scales as $H\propto R^{1/2}$ during infall due to the previous impact of magnetic pressure, leading to $B \propto R^{-1}$.

Not only the strength of the magnetic field, but also its topology evolve significantly as a result of the tidal disruption. In Fig.~\ref{fig:B_lines_geometries} we show snapshots of the stream at time $t = 7.96 \, \rm{h}$ for three different simulations initialized with a uniform (top), toroidal (centre) and tangled (bottom) magnetic field. These panels show that regardless of the initial geometry, eventually the dominant component of the magnetic field is the one parallel to the stream's elongation, as expected from the re-orientation of the magnetic field lines. \footnote{Because of this alignment, magnetic pressure will eventually drive a fast expansion of the stream in all cases. In the uniform toroidal configuration the fraction of magnetic field that is initially aligned with the stretching direction $B_{\parallel,\rm i}$ is similar to the linear field case, and accordingly we find no noticeable difference in the time at which magnetic effects become important. By contrast, in the case of a tangled field the onset of magnetic pressure support is delayed by a factor of a few. This is because the tangled field is inhomogeneous, with its strength decreasing with distance from the stellar center. This reduces the average $B_{\parallel,\rm i}$ within a given stream element, causing $t_{\rm mag}$ to increase relative to the linear and the toroidal configurations.} Aside from this main feature, some differences can be observed between different geometries. Most notably, in the toroidal and tangled cases there are regions along the stream where the longitudinal component of the magnetic field changes sign. These arise from loops in the initial magnetic field structure, which get stretched during the disruption. 

To quantify these rapid changes in the magnetic field we compute its curl throughout the stream using the following SPH formulation of the curl of a vector field \citep[][]{PriceMonaghan2004} for particle $a$
\begin{equation}\label{eq:curl_of_B}
    (\vec{\nabla} \times \vec{B})_{a} = \sum_{b} \frac{m_{b}}{\rho_{b}} (\vec{B}_{a}-\vec{B}_{b}) \times \vec{\nabla}_{a}W_{ab},
\end{equation}
where the sum is over the neighbors and $W_{ab}=(W_{ab}(h_{a})-W_{ab}(h_{b}))/2$, where $W_{ab}(h)$ is the cubic spline and $h$ is the smoothing length. We then slice the stream in thin sections based on their orbital energy parametrized by $\mu$ and compute, using equation~(\ref{eq:curl_of_B}), the average of the vertical component of the curl of the magnetic field over the particles belonging to the slice. In Fig.~\ref{fig:Curl_of_B} we show the ratio between the curl of the magnetic field and its strength as a function of $\mu$ at time $t=7.96 \, \rm{h}$ (that is the same time as Fig.~\ref{fig:B_lines_geometries}), for the three different initial field geometries. In the toroidal and tangled cases (see blue and red lines) the curl of the magnetic field is higher than the uniform case by roughly an order of magnitude. This difference arises because, in the toroidal and tangled cases, the loops in the initial configuration force the magnetic field lines to bend on spatial scales smaller than the stream's transverse size. In the uniform case, instead,  the curl is non zero only due to the longitudinal curvature of the stream itself. 
Furthermore, the toroidal case (see blue line) is roughly symmetric around $\mu=0$, while the tangled one (see red line) shows an asymmetry, which reflects the higher curvature in the unbound part of the stream (see the red arrow in Fig.~\ref{fig:Curl_of_B} and in the bottom panel of Fig.~\ref{fig:B_lines_geometries}, marking the point of highest curvature in the tangled case). The location of high curvature are of interest because this is where the non-ideal MHD effects could take place preferentially.

\subsubsection{Width and magnetic field in the infalling stream}
\begin{figure}
    \hspace{-0.2cm}
    \includegraphics[width=1.05\columnwidth]{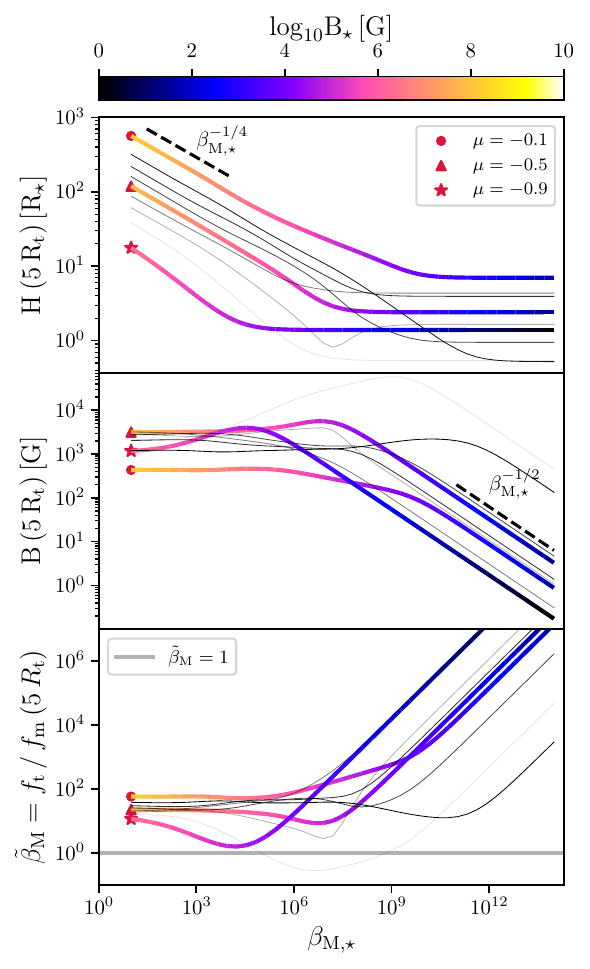}
    \caption{$\tilde{\beta}_{\rm{M}}$ (bottom panel), magnetic field in gauss (centre) and stream width (top) vs initial plasma beta for different infalling stream elements at $5 \, R_{\rm{t}}$, as obtained with the semi-analytic model. The black transparent lines correspond to stream elements with orbital energies between those in the legend, with less bound ones appearing more transparent. The black dashed segments indicate the expected analytic scalings and the grey line in the top panel corresponds to $\tilde{\beta}_{\rm{M}}=1$, meaning that below this line stream elements are dominated by the magnetic field. The colour bar indicates the initial magnetic field.}
    \label{fig:B_vs_beta}
\end{figure}
The semi-analytic model allows us to follow the evolution of all the bound stream elements as these infall towards the black hole, without the inaccuracies of the simulation caused by a lack of resolution. We take advantage of this and show in Fig.~\ref{fig:B_vs_beta} some of the MHD properties of the bound part of the stream: the width, the magnetic field and $\tilde{\beta}_{\rm{M}}$, at a distance of $5 \, R_{\rm{t}}$ as a function of the initial plasma beta, for stream elements with different orbital energies $\mu=-0.1,-0.5,-0.9$ (coloured lines) and other values in between (thin grey lines). These results can help inform the choice of initial conditions for studying the later stages of a TDE, for example through numerical simulations like those by \citet{sadowski2018} and \citet{Meza25}.

The width behaviour (top panel) can be linked to the width at apocentre $H(R_{\rm{apo}})$ shown in Fig.~\ref{fig:H_vs_beta}. This is because, as explained in Section~\ref{Ss:Stream width increase under magnetic pressure}, if the magnetic field becomes dynamically important, the width follows $H\propto R^{1/2}$ during infall, such that $H(5\,R_{\rm t})\propto H(R_{\rm apo}) \propto \beta^{-1/4}_{\rm M,\star}$ for a given $\mu$.

The dependence of the magnetic field strength (central panel) can be understood again from magnetic flux conservation:
\begin{equation}\label{eq:B10Rt}
    B (5 \, R_{\rm{t}}) = B_{\star} \frac{R^{2}_{\star}}{H (5 \, R_{\rm{t}})^{2}} \propto \begin{cases} \beta_{\rm{M},\star}^{-1/2} & \hspace{0.2cm} \mbox{high $\beta_{\rm{M},\star}$} \\ \rm{const} & \hspace{0.2cm} \mbox{low $\beta_{\rm{M},\star}$},
\end{cases}
\end{equation}
where we have used the relation $B_{\star} \propto \beta^{-1/2}_{\rm{M},\star}$ and equation~(\ref{eq:beta-14}). This is consistent with the result in Fig.~\ref{fig:B_vs_beta} for low $\beta_{\rm{M},\star}$, which shows that the magnetic field plateaus at values between $10^2 \, \rm G$ and $10^4\, \rm G$, independent of the stream element's orbital energy. This implies that, as the stream becomes magnetically dominated, the final magnetic field distribution at a given distance from the black hole is independent of the stellar magnetic field. We illustrate how important the magnetic field in the infalling stream is by computing the ratio $\tilde{\beta}_{\rm{M}}=\ddot{H}_{\rm{t}}/\ddot{H}_{\rm{m}}$ (bottom panel) since during the infall the tidal force dominates the dynamics. The result shows that $\tilde{\beta}_{\rm{M}}>1$ for most $\beta_{\rm{M},\star}$ and $\mu$, indicating that the magnetic field is generally not dynamically important at this distance. However, as shown by the grey line reaching $\tilde{\beta}_{\rm M} <1$, it can still be significant during infall in some cases.

\section{Discussion}\label{S:Discussion}
In this section, we discuss the potential impact of some physical processes that for simplicity were neglected in this work. We then describe the implications of our results on the evolution of both the bound and unbound portions of the stream during the later stages of the event.
\subsection{Effect of recombination}
By using an adiabatic equation of state and relying on the ideal-MHD assumption, we have neglected in this work the effect of gas recombination. During a TDE the gas is expected to start recombining when the temperature falls below $T_{\rm{rec}}=10^{4}\,K$, that is, when the stream reaches a distance $R_{\rm{rec}}$ given by
\begin{equation}
    R_{\rm{rec}} = R_{\rm{t}} \left(\frac{T_{\star}}{T_{\rm{rec}}}\right)^{1/2} \approx 30 \, R_{\rm{t}} \left(\frac{T_{\star}}{10^{7}\,K}\right)^{1/2},
\end{equation}
where we have used that $T \propto P/\rho \propto \ R^{-2}$ during the early evolution. This corresponds to a time after disruption of $\approx 3 \, \rm days$. From this stage recombination can inject thermal energy in the stream, which could be either radiated away (this has been proposed to lead to a "recombination transient" \citep{Kasen2010}) or could cause the gas to expand. \citet{Coughlin2023} recently found with semi-analytic calculations that recombination is mostly radiatively inefficient because of the high optical depth of the gas at this point of the evolution. As a result, the main effect of recombination is to cause the stream to expand, an effect also seen in the simulations by \citet{Steinberg2024}. We plan to further investigate the role of recombination during the early stages of a TDE with future numerical simulations. 

If recombination leads to a rapid growth of the stream width, it may dominate over the expansion caused by magnetic fields. The stellar magnetic field is the key parameter that determines which of the two effects becomes dominant first. In parts of the stream that are initially in hydrostatic equilibrium, the magnetic field becomes dynamically important within a distance $R_{\rm rec}=30\,R_{\rm t}$ for an initial plasma beta $\beta_{\rm M}\lesssim 10^4$, which in the core of the star corresponds to $\sim 10^6 \, \rm G$. For larger values of $\beta_{\rm{M}}$, the stream expansion driven by recombination is expected to start earlier than that caused by magnetic pressure support.

\subsection{Impact of non-ideal MHD effects}\label{Ss:non-ideal MHD}
Once the gas recombines, the ideal MHD assumption could break down. As a result, the ideal term in the induction equation\footnote{The induction equation for a partially ionized gas is \citep[][]{PandeyWardle2008}
\begin{equation}
\begin{split}
\frac{\partial \vec{B}}{\partial t} = \nabla \times \Bigg[
    & \vec{v} \times \vec{B}
    - \frac{c^2 m_{\rm e} (\nu_{\rm ei}+\nu_{\rm en})}{4\pi e^2 n_{\rm e}} \nabla \times \vec{B} \\
    & - \frac{c}{4\pi e n_{\rm e}} (\nabla \times \vec{B}) \times \vec{B} \\
    & + \left( \frac{\rho_{\rm n}}{\rho} \right)^2 
      \frac{1}{4\pi \nu_{\rm in} \rho_{\rm i}}
      \big[ (\nabla \times \vec{B}) \times \vec{B} \big] \times \vec{B}
\Bigg].
\end{split}
\end{equation}
}
$\mathcal{I} \sim v_{\rm typ}$, where $v_{\rm{typ}}$ is the typical fluid velocity, may no longer dominate over the diffusion terms associated with each non-ideal MHD effect. These are Ohmic dissipation $\mathcal{O}$, Hall effect $\mathcal{H}$ and ambipolar diffusion $\mathcal{A}$, and they can be approximated as
\begin{equation}\label{eq:OHA}
    \mathcal{O} \approx \frac{c^{2}m_{\rm e}(\nu_{\rm en}+\nu_{\rm ei})}{4\pi e^{2}n_{\rm e}L_{\rm{typ}}},
    \mathcal{H} \approx \frac{c B}{4\pi e n_{\rm e}L_{\rm{typ}}},
    \mathcal{A} \approx \frac{(m_{\rm n} n_{\rm n})^{2}}{4\pi m_{\rm i}n_{\rm i}\nu_{\rm in}L_{\rm{typ}}}\frac{B^{2}}{\rho^{2}},
\end{equation}
where $c$ is the speed of light, $e$ is the electron charge, $m_{\rm j}$ and $n_{\rm j}$ are the mass and number density of particles of species $\rm j$, $\nu_{\rm ij}$ is the collision frequency between particles of species $\rm i$ and $\rm j$. These terms have the same dimension of the ideal one $\mathcal{I}$, namely a velocity, and are obtained by approximating $|\nabla \times \vec{B}| \approx B /L_{\rm typ}$ in the induction equation, where $L_{\rm typ}$ is the typical length scale over which the magnetic field changes.

In the initial star the gas is fully ionized, such that $n_{\rm n}\ll n_{\rm{i}}=n_{\rm e}$ and $\nu_{\rm en}\ll \nu_{\rm ei}$. Furthermore, the electron-ion and the ion-neutral collision frequencies are given, respectively, by $\nu_{\rm ei}=51\, (n_{\rm e}/\rm{cm^{-3}})\,(T/\rm{K})^{-3/2} \, \rm{s^{-1}}$ \citep[][]{PandeyWardle2008}, and $\nu_{\rm in}= \langle \sigma v \rangle_{\rm in} \, \rho_{\rm n}/(m_{\rm n}+m_{\rm i})$, where $\langle \sigma v \rangle_{\rm in}=1.9 \times 10^{-9} \, \rm{cm^{3}s^{-1}}$ is the ion-neutral momentum transfer rate coefficient \citep[][]{Draine1983}. This can be used to estimate the ratios between the non-ideal and the ideal MHD terms in the initial star:  
\begin{align}\label{OI_star}
    \left(\frac{\mathcal{O}}{\mathcal{I}}\right)_{\star} \approx 1.5 \times 10^{-16} \left(\frac{T_{\star}}{10^7 \, \rm{K}}\right)^{-3/2} \left(\frac{M_{\star}}{M_{\odot}}\right)^{-1/2}\left(\frac{R_{\star}}{R_{\odot}}\right)^{-1/2},
\end{align}
\begin{align}\label{HI_star}
    \left(\frac{\mathcal{H}}{\mathcal{I}}\right)_{\star} \approx 4.6 \times 10^{-25} \left(\frac{B_{\star}}{1\,G}\right)\left(\frac{M_{\star}}{M_{\odot}}\right)^{-3/2}\left(\frac{R_{\star}}{R_{\odot}}\right)^{5/2},
\end{align}
\begin{align}\label{AI_star}
    \left(\frac{\mathcal{A}}{\mathcal{I}}\right)_{\star} \approx 1.3 \times 10^{-46} \left(\frac{n_{\rm{n}}/n_{\rm e}}{10^{-10}}\right) \left(\frac{B_{\star}}{1\,G}\right)^{2}\left(\frac{M_{\star}}{M_{\odot}}\right)^{-5/2}\left(\frac{R_{\star}}{R_{\odot}}\right)^{11/2},
\end{align}
where $\rho_{\star}=M_{\star}/R^{3}_{\star}$ and $T_{\star}=10^7 \, \rm{K}$ are approximately the initial density and temperature, and $B_{\star}$ is the stellar magnetic field. As expected, the non-ideal MHD terms are initially all negligible compared to the ideal one. 

Nevertheless, soon after disruption, while the gas is fully ionized, along the transverse size of the stream these ratios all grow following
\begin{equation}
    \left(\frac{\mathcal{O}}{\mathcal{I}}\right) \propto T^{-3/2} H^{-1} \dot{H}^{-1} \propto R^{7/2}
\end{equation}
\begin{equation}
    \left(\frac{\mathcal{H}}{\mathcal{I}}\right) \propto \frac{B}{\rho \, H \, \dot{H}} \propto R^{5/2}
\end{equation}
\begin{equation}
    \left(\frac{\mathcal{A}}{\mathcal{I}}\right) \propto \frac{B^{2}}{\rho^{2} \, H \, \dot{H}} \propto R^{9/2}.
\end{equation}
These are obtained setting $L_{\rm{typ}}=H$ and $v_{\rm{typ}}=\dot{H}$ and by using the scalings $\ell \propto R^{2}$, $H \propto R^{1/2}$ and $t\propto R^{3/2}$, which are valid at early times.
As a result, by the time the stream reaches a distance $R_{\rm{rec}}$, the ratio $\mathcal{O}/\mathcal{I} \approx 2.2 \times 10^{-11}$ is much larger than $\mathcal{H}/\mathcal{I} \approx 2.3 \times 10^{-21} (B_{\star}/1\,G)$ and $\mathcal{A}/\mathcal{I} \approx 5.9 \times 10^{-40} (n_{\rm{n}}/10^{-10}n_{\rm{e}}) (B_{\star}/1\,G)^{2}$. This means that, unless the stellar magnetic field is stronger than $\sim 10^{10} \, G$, Ohmic dissipation will be dominant out of the three non-ideal MHD effects when the stream starts to recombine. 

At distance $R_{\rm{rec}}$ the ionization fraction $x=n_{\rm e}/n_{\rm n}$ in the stream starts to drop. There is a critical value $x_{\rm{crit}}$ below which non-ideal MHD effects become comparable to the ideal one. In the case of Ohmic dissipation this is found by setting $\mathcal{O} = \mathcal{I}$, which gives
\begin{equation}\label{eq:x_ohm}
    x^{\rm{Ohm}}_{\rm{crit}} \approx \frac{m_{\rm e}c^2 <\sigma v>_{\rm en}}{4\pi e^2 H \dot{H}},
\end{equation}
where we have used $\nu_{\rm en}=\langle \sigma v \rangle_{\rm en} \, \rho_{\rm n}/(m_{\rm n}+m_{\rm e})$ and $\langle \sigma v \rangle_{\rm en}=8.28 \times 10^{-10} \, (T/\rm K)^{1/2}\, \rm{cm^{3}s^{-1}}$, that is the electron-neutral momentum transfer rate \citep[][]{Draine1983}. To approximate $x^{\rm Ohm}_{\rm crit}$ we have used the fact that $\nu_{\rm ei}$ becomes negligible compared to $\nu_{\rm en}$ as the gas becomes weakly ionized. Using equation~(\ref{eq:x_ohm}), at $R=R_{\rm{rec}}$\footnote{Using $H \propto R^{1/2}$, $H(R_{\rm rec})=R_{\star}\sqrt{R_{\rm rec}/R_{\rm t}}\approx \sqrt{30} \, R_{\star}$ and using $\dot{H}\approx H/t\propto R^{-1}$, $\dot{H} (R_{\rm rec})\approx \sqrt{GM_{\star}/R_{\star}}/30$.}, the Ohmic critical ionization fraction is $x^{\rm{Ohm}}_{\rm{crit}} \approx 4.2 \times 10^{-14}$. Assuming Ohmic dissipation to remain the dominant term during recombination, once the ionization fraction falls below this critical value $x^{\rm{Ohm}}_{\rm{crit}}$, non-ideal MHD effects cannot be ignored and there might be deviations from the results presented in Section~\ref{S:results}. A more accurate study of the evolution of the ionization fraction in the stream and its effect on the impact of magnetic fields will be the focus of future work.

The above calculations neglect the impact of magnetic field geometry. As shown in Section~\ref{Magnetic field evolution along the stream}, the magnetic field may vary over $L_{\rm typ}<H$ in regions of curved magnetic field, causing the critical ionization fraction to increase and therefore anticipate the emergence of non-ideal MHD effects. For instance, we anticipate that such effects could manifest themselves through magnetic reconnection events happening at the interface between field lines of opposite directions, which we find to naturally form for a initially toroidal and tangled stellar magnetic field (see two lowermost panels in Fig.~\ref{fig:B_lines_geometries}). Since compression of the field lines is required for reconnection to occur, we expect this process to become particularly relevant during the return of the stream to pericentre.

\subsection{Role of magnetic fields in the later evolution of the TDE}
In this work, we studied the dynamical impact of magnetic fields during the evolution of the debris stream around the black hole. Shortly after the stream returns to pericentre, the magnetic field can again become dynamically important. This is because near the black hole the gas is subject to a strong vertical compression that is expected to amplify the magnetic field, potentially making magnetic pressure significant compared to gas pressure. Assuming magnetic flux conservation, this occurs if the stellar magnetic field is greater than $\sim 400 \, \rm{G}$ \citep[][]{BonnerotLu2022}. The effect of this additional magnetic pressure support on the later stages of the TDE remains uncertain, and we plan to investigate it in the future by means of MHD simulations of the nozzle shock. 

Since the nozzle shock is not expected to significantly influence the evolution of the stream thickness \citep[][]{BonnerotLu2022,Hu2025,Kubli2025}, the width at apocentre $H(R_{\rm{apo}})$ shown in Fig.~\ref{fig:H_vs_beta} can be used as an approximation of the width of the stream at the self-intersection point\footnote{For strongly precessing events, the collision happens close to pericenter. However, for the typical parameters considered in this work, $M_{\rm h}=10^{6}\,M_{\odot}$, $M_{\star}=M_{\odot}$ and $\beta=1$, the intersection point is closer to apocenter \citep[][]{BonnerotStone2021}.}. Here, simulations suggest that the shock resulting from the stream-stream collision cannot account for the bulk of the observed emission \citep[][]{Huang_2024,BonnerotLu2020}. This is because of adiabatic losses that limit the emerging luminosity to $L \approx 3 \times 10^{42} (H/10\,R_{\odot})^{2/3} \rm{erg \, s^{-1}}$, where $H$ is the stream width \citep[][]{BonnerotStone2021}. Stream thickening due to magnetic pressure can make the gas more optically thin, leading to a larger emitted luminosity of $\sim 10^{43} - 10^{44} \rm{erg \, s^{-1}}$,  thereby contributing more to the electromagnetic signatures.\footnote{An additional source of energy dissipation may be provided by magnetic reconnection, which is expected to occur as the two streams, threaded by longitudinal magnetic field lines, collide at an angle as set by the geometry of the problem.}

If the black hole spins, the orbital plane of the stream can change because of the Lense-Thirring effect, which may prevent the stream from self-intersecting. An increased thickness due to the dynamical impact of magnetic fields, makes it more likely for the collision to occur despite this relativistic effect. Furthermore, the top panel of Fig.~\ref{fig:B_vs_beta} suggests that there may be extreme scenarios where the thickness of the stream becomes similar to the distance from the black hole close to pericentre, which would lead to qualitative differences in the following evolution, perhaps causing the stream-stream interaction to occur at a wide range of distances from the black hole.

Not only the stream transverse structure, but also the magnetic field evolution at early times is crucial to the understanding of the later phases of a TDE, as this can influence the accretion disk evolution. The magnetic field in the stream is expected to be advected by the quasi-spherical outflow resulting from the self-crossing shock, which will boost the out-of-plane magnetic field component, potentially creating preferentially poloidal field lines in the formed disk. This may not occur if the collision is less powerful, for example, because of nodal precession, in which case it is possible that the magnetic field in the disk remains toroidal. This orientation of the magnetic field can affect the degree of stabilization of the disk against thermal and viscous instabilities. \citet{Mishra2022} found with numerical simulations that strong vertical magnetic fields are more effective at supporting the disk. Furthermore, the poloidal component of the magnetic field is crucial to the development and sustainment of the MRI in the accretion disk formed following the collision. Indeed, linear \citep[e.g.][]{BalbusHawley1992} and non-linear \citep[e.g.][]{Hawley1995, BalbusHawley1998} studies of the MRI have shown that this instability grows fastest and produces a stronger sustained turbulence
when the magnetic field has a poloidal component. 

This work provides self-consistent initial conditions for studying disk formation with the inclusion of magnetic fields. This will enable investigations of the onset of the MRI during a TDE, thereby shedding light on whether the luminosity arising from the accretion disk is powered by magnetic processes. Current numerical MHD simulations of TDE accretion disks suggest that the MRI does not play a significant role in driving accretion \citep[][]{Curd2021,sadowski2018,Meza25}. However, this result may be influenced by low resolution and by the choice of initial conditions in these simulations, particularly the neglect of the early evolution of the debris stream around the black hole.  

The dynamical impact of magnetic fields studied in this work has implications also for the unbound tail of the stream, particularly in connection with the radio emission observed in many TDEs. The interaction between the unbound part of the stream and the circumnuclear medium has been proposed as a mechanism to produce synchrotron radiation through shock interactions \citep[][]{Yalinewich2019}. However, this effect is typically not favoured to explain the observed radio emission due to the small cross section of the unbound stream compared to other types of outflow expected to form during a TDE. The increase in the magnetized stream thickness we find in this work could enhance the level of radiation emission compared to these previous estimates. As described in Section~\ref{Ss:Stream width increase under magnetic pressure}, when magnetic pressure becomes important, for some time the width is expected to follow $H \propto R^{5/4}$ (see Fig.~\ref{fig:H_vs_R}). As a result, the opening angle of the stream $\theta\approx 2\arctan{(H/R)}$ increases with the distance from the black hole. To widen the angle by a factor $ 10$, the stream must reach a distance $R=10\,R_{\rm t}\beta^{1/3}_{\rm M,\star}$. For a stellar magnetic field of $1\,\rm G$, $\beta_{\rm M,\star}\approx 10^{16}$, yielding $R \approx 5\, \rm{pc}$. As the radio luminosity is proportional to the solid angle \citep[e.g.][]{LuBonnerot2019}, it may be enhanced due to this effect, making it a more promising mechanism for the radio emission observed in TDEs. 

Furthermore, during the interaction between the stream and the ambient medium, magnetic fields may sustain the stream against the Kelvin-Helmoltz instability, which would otherwise lead to its dissolution in the ambient medium \citep[][]{Bonnerot2016,Berlok2019}. Similarly, magnetic fields may protect the bound part of the stream from shock interactions with the disk formed during the TDE \citep[][]{Steinberg2024,Huang_2024}, or with a pre-existing disk. These disks can disrupt the bound debris stream, ultimately stalling its fallback towards the black hole and thereby suppressing the TDE luminosity, an effect that has been proposed to explain the abrupt cutoff observed in the light curve of some TDEs \citep[][]{Kathirgamaraju2017}. One of the parameters that could determine the presence or absence of such cutoff is the magnetic field in the stream, which could inhibit its dissolution by the disk \citep[][]{McCourt2015,Berlok2019}, potentially delaying the gas energy dissipation caused by these shock interactions and its subsequent circularization.

By contrast, magnetic fields could destabilize the stream if the MRI were to develop within it, potentially leading to its disruption through turbulence. The closest work related to this problem is that by \citet{Chan_2024}, who study analytically the development of the MRI in eccentric disks. A future generalization to the parabolic case would enable the investigation of MRI development in a TDE stream.

\section{Conclusions}\label{S:Conclusions}
We studied the tidal disruption of a magnetized star by a supermassive black hole, following the stream evolution up until its return near the black hole. We employed numerical simulations whenever feasible, and adopted the semi-analytic approach when the simulations were computationally prohibitive or to allow for a broader exploration of the parameter space. The main results and conclusions from this study can be summarized as follows:

\begin{enumerate}[label=(\roman*), leftmargin=*, align=left]

    \item As previously found by \citet{Guillochon2017} and \citet{Bonnerot2017}, magnetic fields can eventually become dynamically important along the transverse direction of the stream during the early stages of a TDE. We find through analytic calculations (see Section~\ref{ss:Onset of magnetic pressure support}) that the onset of this effect and the conditions under which it occurs vary significantly along the stream. While in most of the unbound mass magnetic fields can become significant at any time after disruption, bound stream elements are limited by their return to the black hole, which sets a lower limit on the required stellar magnetic field. We use these calculations to estimate that at least $50\,\%$ of the mass of the stream becomes magnetically supported within $\sim 1 \, \rm{yr}$ after disruption, if the stellar magnetic field is above $\sim 10^4 \, \rm{G}$ (see Fig.~\ref{fig:mass_fraction}).\\
    
    \item Our simulations are the first to demonstrate that during the stream evolution around the black hole the stellar magnetic field can drive a fast increase in its width. Notably, we find that throughout the stream the predominance of magnetic pressure is accompanied by a phase of transverse equilibrium between magnetic and tidal force, which causes the stream thickness to quickly grow following $H \propto R^{5/4}$ (see Fig.~\ref{fig:H_vs_R}). This ends the stream's confinement by self-gravity in high density regions that are initially in hydrostatic equilibrium, while in the more dilute gas that is initially dominated by the tidal force, this rapid growth is even faster than that caused by homologous expansion.\\
    
    \item We find, with both numerical simulations and semi-analytic modelling, that if the magnetic field becomes dynamically important in the bound part of the stream, during its infall towards the black hole the thickness increases with decreasing plasma-beta following a simple power law $H \propto \beta^{-1/4}_{\rm M,\star}$ (see Fig.~\ref{fig:H_vs_beta}). As a result, depending on the initial magnetic field strength, the stream width can deviate from the purely hydrodynamic evolution by a factor ranging from a few up to one or two orders of magnitude in the most magnetized cases.\\ 

    \item The magnetic properties of the returning stream vary with the strength and topology of the stellar magnetic field. For weak initial fields, as the stream approaches the black hole, the stream thickness, and thus its magnetic field, remain similar to their initial values. If instead the stellar magnetic field is strong enough to become dynamically important, the magnetic field of the returning stream near the black hole is between $\sim 10^2$ to $\sim10^4 \, \rm G$, regardless of its initial value (see Fig.~\ref{fig:B_vs_beta}). As found in previous works, the disruption makes the magnetic field mostly aligned with the stream elongation. Furthermore, in our numerical simulations we find that initializing the star with a purely toroidal or a tangled magnetic field produces regions of stronger magnetic field curvature and field reversal across the stream compared to the uniform field case (see Fig.~\ref{fig:B_lines_geometries}).\\

    \item A few days after disruption, when the stream reaches a distance of $R\approx 30\,R_{\rm{t}}$ from the black hole, the gas is expected to recombine into neutral atoms as a result of adiabatic cooling. We estimated that Ohmic dissipation is the first non-ideal MHD effect to develop, and it may cause deviations from the results presented in this work if during recombination the ionization fraction falls below $x=n_{e}/n_{\rm n}\approx 10^{-14}$. These departures from ideal MHD may be further enhanced in regions of strong magnetic field gradients.\\ 

    \item If the outflow from the stream–stream collision is approximately spherical, we expect the magnetic field in the disk to develop a poloidal component because of magnetic flux conservation. Conversely, an asymmetric collision that produces an outflow mostly in the orbital plane may cause the magnetic field to remain confined to the plane, as in the stream. This orientation of the magnetic field can have important consequences for the late evolution of TDEs, influencing the efficiency of MRI development in the disk and the formation of jets.\\

    \item In the unbound part of the stream, the fast increase in width, $H \propto R^{5/4}$, causes the opening angle of the stream to grow over time, potentially enhancing the luminosity produced during the interaction with the ambient medium. This finding suggests that magnetic fields may enhance the contribution of the unbound debris to the radio emission observed from TDEs. 
\end{enumerate}

This work was strongly motivated by the need for future investigations of the later stages of a TDE that incorporate magnetic fields. Such studies will be crucial to clarify the influence of magnetic fields on the observational signatures of TDEs, particularly the X-ray and radio emission and the formation of relativistic jets. 

\section*{Acknowledgments}

We thank Eric Coughlin, Emily Hatt, Hui Li, Greg Salvesen, Re'em Sari and Nicholas Stone for useful discussions. Funded by the European Union (ERC, Unleash-TDEs, project number 101163093). Views and opinions expressed are however those of the author(s) only and do not necessarily reflect those of the European Union or the European Research Council. Neither the European Union nor the granting authority can be held responsible for them. This research was supported in part by grant NSF PHY-2309135 to the Kavli Institute for Theoretical Physics (KITP). The research leading to this work received funding from the Independent Research Fund Denmark via grant ID 10.46540/3103-00205B. The computations described in this paper were performed using the University of Birmingham's BlueBEAR HPC service, which provides a High Performance Computing service to the University's research community. We acknowledge the use of SPLASH \citep[][]{Price2007} for the visualization of the results. 

\section*{Data Availability}
The data underlying this paper will be shared on reasonable request to the corresponding author. A public version of the GIZMO code is available at \url{http://www.tapir.caltech.edu/~phopkins/Site/GIZMO.html}.



\bibliographystyle{mnras}
\bibliography{bibliography} 




\appendix

\section{Magnetic acceleration equations}\label{app:Magnetic acceleration equations}
The magnetic acceleration $\ddot{\textbf{r}}_{\rm{mag}}$ can be written as: 

\begin{equation}
    \ddot{\textbf{r}}_{\rm{mag}}=-\frac{\nabla B^{2}}{8\pi}+\frac{1}{4\pi}(\textbf{B}\cdot \nabla)\textbf{B}.
    \label{eqA:ddot_rm}
\end{equation}
In the framework of the semi-analytic model presented in Section~\ref{sec:Semi-analytic model of a magnetized TDE}, we have that:
\begin{equation}
    \textbf{B} = B_{\parallel} \textbf{e}_{\parallel} + B_{\perp} \textbf{e}_{\perp} + B_{\rm{z}} \textbf{e}_{\rm{z}},
\end{equation}
\begin{equation}
    \nabla = \frac{\partial }{\partial \ell}\textbf{e}_{\parallel} + \frac{\partial }{\partial \Delta} \textbf{e}_{\perp} + \frac{\partial }{\partial H} \textbf{e}_{\rm{z}},
\end{equation}
and
\begin{equation}
    \ddot{\textbf{r}}_{\rm{mag}} = \ddot{\ell}_{\rm m} \,  \textbf{e}_{\parallel} + \ddot{\Delta}_{\rm m} \, \textbf{e}_{\perp} + \ddot{H}_{\rm m} \,  \textbf{e}_{\rm{z}},
\end{equation}
where $\textbf{e}_{\parallel}$,$\textbf{e}_{\perp}$,$\textbf{e}_{\rm z}$ are orthogonal unit vectors in a frame that co-rotates with the center of mass of a section. By projecting $\ddot{\textbf{r}}_{\rm{mag}}$ along the vertical direction we obtain the following equation for the vertical component of the magnetic acceleration: 
\begin{equation}\label{eq:ddot_Hm_partial}
    \ddot{H}_{\rm m} = -\frac{\partial}{\partial H}\left(\frac{B^2_{\parallel}}{8\pi}\right) -\frac{\partial}{\partial H}\left(\frac{B^2_{\perp}}{8\pi}\right) + \left(\frac{B_{\perp}}{4\pi}\right) \frac{\partial B_{\rm z}}{\partial \Delta} + \left(\frac{B_{\parallel}}{4\pi}\right) \frac{\partial B_{\rm z}}{\partial \ell}.
\end{equation}
Assuming that the magnetic field gradient across the transverse size of the stream is much larger than the gradient along the stream elongation, the last term on the right-hand side of equation~(\ref{eq:ddot_Hm_partial}) is much smaller than the first term ($\partial B_{\rm z}/\partial \ell \ll \partial B_{\parallel}/\partial H$) and therefore it can be neglected. Furthermore, by approximating $\partial/\partial H = - 1/H$ and $\partial/\partial \Delta = - 1/\Delta$, equation~(\ref{eq:ddot_Hm_partial}) simplifies to
\begin{equation}\label{eq:ddot_Hm_appendix}
    \ddot{H}_{\rm m} = \frac{1}{4\pi}\left(\frac{B^2_{\parallel}}{2 H} +\frac{B^2_{\perp}}{2 H} - \frac{B_{\perp} B_{\rm z}}{\Delta}\right).
\end{equation}
In a similar way, we approximate the equations for the magnetic acceleration of the in-plane width and along the stream elongation as follows
\begin{equation}\label{eq:ddot_Dm_appendix}
    \ddot{\Delta}_{\rm m} = \frac{1}{4\pi}\left(\frac{B^2_{\parallel}}{2 \Delta} +\frac{B^2_{\rm z}}{2 \Delta} - \frac{B_{\perp} B_{\rm z}}{ H}\right),
\end{equation}
\begin{equation}\label{eq:ddot_ellm_appendix}
    \ddot{\ell}_{\rm m} = -\frac{1}{4\pi}\left(\frac{B_{\perp} B_{\parallel}}{ \Delta}+\frac{B_{\rm z} B_{\parallel}}{ H}\right).
\end{equation}
Despite the above approximations, equations~(\ref{eq:ddot_Hm_appendix}),~(\ref{eq:ddot_Dm_appendix}) and~(\ref{eq:ddot_ellm_appendix}) guarantee that $\ddot{\textbf{r}}_{\rm mag} \cdot \textbf{B}=0$, meaning that there is no magnetic force along magnetic field lines, as physically expected. Furthermore, we expect $\ddot{\ell}_{\rm m}$ to never become important during the stream evolution around the black hole. This is because, at early times, before the magnetic field aligns with the stream elongation, $\ddot{\ell}_{\rm m}$ is subdominant compared to the acceleration associated with the tidal force. Later, when the magnetic field aligns with the stream elongation, the magnetic acceleration can dominate the dynamics, however, by this time, both $B_{\perp}$ and $B_{\rm z}$ are effectively negligible so that $\ddot{\ell}_{\rm m}=0$. Because of this, we only include equations~(\ref{eq:ddot_Hm_appendix}) and~(\ref{eq:ddot_Dm_appendix}) in the dynamical equations that are solved in our semi-analytic model.

\bsp	
\label{lastpage}

\end{document}